\documentclass[12pt]{article}

\usepackage{amsmath}
\usepackage{amsthm}
\usepackage{amssymb}
\usepackage{graphicx}
\usepackage{psfrag}

\makeatletter
\def\fixNumberingInArticle{
\@addtoreset{equation}{section}
\renewcommand{\theequation}{\thesection.\arabic{equation}}  
}

\def\fixNumberingInAppendix{
\setcounter{section}{0}
\renewcommand{\theequation}{\Alph{section}.\arabic{equation}}
\renewcommand{\thesection}{\Alph{section}}
\setcounter{equation}{0}
}
\makeatother

\newcommand{\Am}{A_{\freq,0}}

\newcommand{\RII}[1]{R_{\sol}^{(2)}(#1)}
\newcommand{\RIII}[1]{R_{\sol}^{(3)}(#1)}
\newcommand{\NII}[1]{N_{\sol}(#1)}
\newcommand{\pRIII}[1]{{R_{\sol}^{(3)}}'(#1)}

\newcommand{\RotSpace}{\mathrm{SO}}
\newcommand{\dotp}[2]{\langle #1,#2\rangle}

\newcommand{\Mm}{n}
\newcommand{\Mt}{\tilde{n}}

\newcommand{\freq}{\mu}
\newcommand{\om}{\Omega_{\sol}}
\newcommand{\iom}{\Omega_{\sol}^{-1}}
\newcommand{\piom}{\Omega_{\sol_{\Par}}^{-1}}
\newcommand{\Miom}{\left. \iom\right|_{\{z_{k}\}}}

\newcommand{\J}{J}

\newcommand{\bone}{\mathchoice {\rm 1\mskip-4mu l} {\rm 1\mskip-4mu l}
{\rm 1\mskip-4.5mu l} {\rm 1\mskip-5mu l}}
\newcommand{\Ham}{\mathcal{H}}

\newcommand{\sol}{\eta}
\newcommand{\solp}{\sol_\sigma}
\newcommand{\solw}{\sol_\freq}

\newcommand{\stfrm}{\omega}
\newcommand{\srfrm}{\stfrm_{\sol_{\Par}}}
\newcommand{\tfrm}[2]{\stfrm(#1,#2)}

\newcommand{\rfrm}[2]{\srfrm(#1,#2)}

\newcommand{\cf}{{\it cf.\/}\ }

\newcommand{\vip}{v}

\newcommand{\set}[1]{\mathrm{#1}}

\newcommand{\Hess}{\mathop{\mathrm{Hess}}}

\renewcommand{\exp}[1]{\mathrm{e}^{#1}}
\newcommand{\RE}{\mathop{\set{Re}}}
\newcommand{\IM}{\mathop{\set{Im}}}
\renewcommand{\i}{{\rm i}}

\newcommand{\diag}{\mathop{\mathrm{diag}}}

\newcommand{\Range}[1]{\set{Ran}( #1 )}
\newcommand{\Null}[1]{\set{N}( #1 )}

\newcommand{\adj}{*}
\newcommand{\adjoint}[1]{#1^{\adj}}

\newcommand{\Eq}{Eq.\/~}
\newcommand{\eq}{eq.\/~}
\newcommand{\Eref}[1]{\Eq\eqref{#1}}
\newcommand{\eref}[1]{\eq\eqref{#1}}

\newcommand{\ie}{{\it i.e.\/}, }

\newcommand{\eg}{{\it e.g.\/}, }
\newcommand{\diff}{\mathop{\mathrm{\mathstrut{d}}}\!}

\newcommand{\Hone}{\set{H}_1}
\newcommand{\Htwo}{\set{H}_2}
\newcommand{\Sob}[1]{\set{H}_{#1}}
\newcommand{\C}[1]{\set{C}^{#1}}
\newcommand{\Ltwo}{\set{L}^2}
\newcommand{\Lp}[1]{\set{L}^{#1}}

\newcommand{\sind}[1]{{\text{{\tiny $#1$}}}}
\newcommand{\eps}{{\epsilon_{\sind{V}}}}
\newcommand{\epsE}{{\epsilon_{\sind{0}}}}
\newtheorem{proposition}{Proposition}
\newtheorem{step}{Step}
\newtheorem{lemma}{Lemma}
\newtheorem{remark}{Remark}

\newtheorem{theorem}{Theorem}
\newtheorem{corollary}{Corollary}[theorem]

\newcommand{\Sn}{\mathcal{S}}

\newcommand{\En}{\mathcal{E}}
\newcommand{\Nn}{\mathcal{N}}
\newcommand{\Ew}{\En_{\freq}}

\newcommand{\DE}{\En'}
\newcommand{\LL}{\mathcal{L}_\sol}
\newcommand{\LLm}{\mathcal{L}_{\sol_{\freq}}}
\newcommand{\Le}{\mathcal{L}_{\sol,\sigma}}
\newcommand{\Mf}{\set{M}_{\mathrm{s}}}
\newcommand{\TM}{\set{T}_{\sol}\Mf}
\newcommand{\Span}{\mathop{\mathrm{span}}}
\newcommand{\Oh}{\mathcal{O}}
\newcommand{\nrmHo}[1]{\|#1\|_{\Hone}}

\newcommand{\nrmL}[1]{\|#1\|_{\Ltwo}}
\newcommand{\nrmD}[1]{\|#1\|_{\Sob{-1}}}

\newcommand{\Laplace}{\Delta}

\newcommand{\zvec}{z}
\newcommand{\za}{\zvec_{\mathrm{t}}}
\newcommand{\zg}{\zvec_{\mathrm{g}}}
\newcommand{\zt}{\zvec_{\mathrm{b}}}
\newcommand{\zn}{\zvec_{\mathrm{s}}}

\newcommand{\tr}{\za}
\newcommand{\gu}{\zg}
\newcommand{\ga}{\zt}
\newcommand{\sa}{\zn}

\newcommand{\Pro}{\mathcal{P}_{\sol_{\Par}}}

\newcommand{\nrm}[1]{\|#1\|}
\newcommand{\rdot}[2]{\langle #1,#2 \rangle}
\newcommand{\gN}{\set{N}_{\mathrm{gen}}}
\newcommand{\TB}{\mathcal{S}_{a \vip \gamma}}

\newcommand{\Tb}{\hat{\mathcal{S}}_{a \vip \gamma \freq}}

\newcommand{\mm}{m}
\newcommand{\mw}{\mm(\freq)}
\newcommand{\mmp}{\mm'}

\newcommand{\VR}{\mathcal{R}_{V}}

\newcommand{\WW}{|\|w\||_{\Hone}}

\newcommand{\setPar}{\Sigma}
\newcommand{\Par}{\sigma}
\newcommand{\spow}{s}

\newcommand{\Anrm}[1]{|#1|_{\infty}}
\newcommand{\anrm}[1]{|#1|}

\newcommand{\lk}{\mathcal{K}}
\newcommand{\lss}{\underline{\lk}}
\newcommand{\Ass}{\underline{\alpha}}
\newcommand{\Lal}{\alpha\cdot \lk}
\newcommand{\Sal}{\Ass\cdot \lss}

\newcommand{\Rot}{\mathcal{T}^{\textrm{g}}_{\gamma}}
\newcommand{\Tr}{\mathcal{T}_a^{\textrm{tr}}}
\newcommand{\Bo}{\mathcal{T}_{\vip}^{\textrm{gal}}}
\newcommand{\Sc}{\mathcal{T}_{\freq}^{\textrm{s}}}
\newcommand{\Ph}{\mathcal{T}_{\vip}^{\textrm{b}}}
\newcommand{\SRot}{\mathcal{T}^{\textrm{r}}_{R}}
\newcommand{\CC}{\mathcal{T}^{\textrm{c}}}

\newcommand{\Man}{\set{M}}
\newcommand{\TMan}[1]{\set{T}_{#1}\set{M}}
\newcommand{\Ip}[1]{I_{#1}}
\newcommand{\dIp}[1]{\dot{I}_{#1}}

\fixNumberingInArticle
\title{Solitary wave dynamics in an external potential}
\author{J. Fr\"{o}hlich$^1$, S. Gustafson$^{2}$
\thanks{supported by NSERC grant 22R80976},
B. L. G. Jonsson$^{3,4,}$\thanks{the support of Wenner-Gren
Foundation is gratefully acknowledged}, 
I. M. Sigal$^{3,5,}$\thanks{Supported by NSERC under NA7601}\\ {\small
$^1$Institute f\"{u}r Theoretische Physik, ETH H\"{o}nggerberg,
Z\"{u}rich, Schwitzerland.} \\ {\small $^2$Department of Mathematics,
University of British Columbia, Vancouver, Canada.}  \\ {\small 
$^3$Department of Mathematics, University of Toronto, Toronto,
Canada.}\\ {\small $^4$The Alfv\'enlaboratory, Royal Institute of 
Technology, Stockholm, Sweden.}\\  
{\small $^5$Department of Mathematics, University of Notre Dame, 
Notre Dame, IN, USA.}} 

\date{\today}

\begin{document}

\maketitle

\begin{abstract}
We study the behavior of solitary-wave solutions of some generalized
nonlinear Schr\"odinger equations with an external potential. The
equations have the feature that in the absence of the external potential,
they have solutions describing inertial motions of stable solitary waves.

We construct solutions of the equations with a non-vanishing external
potential corresponding to initial conditions close to one of these
solitary wave solutions 
and show that, over a large interval of time, they describe a
solitary wave whose center of mass motion is a solution of Newton's equations
of motion for a point particle in the given external potential, up to
small corrections corresponding to radiation damping.
\end{abstract}


\section{Introduction}

\label{sec:intro}

In this paper we study the effective dynamics of solitary wave
solutions of a class of generalized nonlinear Schr\"{o}dinger
equations with an external potential. These equations have the form
\begin{equation}
   \i \partial_t \psi = (-\Laplace + V)\psi - 
f(\psi) \; ,
   \label{eq:sp1}
\end{equation}
where $\psi:\mathbb{R}^{d}\times \mathbb{R}\mapsto \mathbb{C}$, 
$x$ denotes a point in space $\mathbb{R}^d$, $t\in \mathbb{R}$ 
is time, $\partial_{t} = \frac{\partial}{\partial t}$,
$\Laplace=\sum_{j=1}^d \frac{\partial^2}{\partial x_j^2}$ is the
spatial Laplacian, $V(x)$ is the external potential and is a 
real-valued, bounded slowly varying function on $\mathbb{R}^d$, and 
$f$ is a map from the complex Sobolev space
$\Hone(\mathbb{R}^{d},\mathbb{C})$
to 
$\Sob{-1}(\mathbb{R}^{d},\mathbb{C})$ such that $f(0)=0$, and 
$f(\bar{\psi})=\overline{f(\psi)}$,
describing a nonlinear ``self-interaction''.
Precise assumptions on $V$ and $f$ will be 
given in Section~\ref{sec:res}.  Examples of 
nonlinearities of interest include local nonlinearities
such as
\begin{equation}
f(\psi)=\lambda |\psi|^{2s}\psi, \ \ 0<s<\frac{2}{d},\ \lambda>0,
\label{eq:lnl}
\end{equation}
and Hartree-type nonlinearities
\begin{equation}
f(\psi) = \lambda (W*|\psi|^2)\psi,\ \ \lambda>0, \label{eq:nlnl}
\end{equation}
where $W$ is of positive type,
continuous, spherically symmetric potential function which tends to 0,
as $|x|\rightarrow \infty$ and $W*g(x):=\int W(x-y)g(y)\diff^d y$, 
denotes (spatial) convolution. Of course,  
$\lambda$ can be scaled out by rescaling $\psi$.

We assume that the nonlinearity in \eqref{eq:sp1} is such that the
Cauchy problem has a unique, global solution, $\psi(x,t)$, in
the space 
$\set{C}(\mathbb{R}^+;\Hone(\mathbb{R}^d,\mathbb{C}))\cap 
\C{1}(\mathbb{R}^+;\Sob{-1}(\mathbb{R}^d,\mathbb{C}))$, 
given an initial condition 
$\psi(x,0)=\psi_0(x)\in\Hone(\mathbb{R}^d,\mathbb{C})$. 
Results on the Cauchy problem associated with
\eqref{eq:sp1} can be found in 
\cite{Ginibre+Velo1979,Kato1987,Cazenave1996} (see Section~\ref{sec:res} for
a discussion).

For $V\equiv 0$, eq.~\eqref{eq:sp1} is the usual generalized nonlinear
Schr\"odinger (or Hartree) equation. For self-focusing nonlinearities
(as in example \eqref{eq:lnl} or \eqref{eq:nlnl} with 
$W$ of positive type and $\lambda>0$ large enough), it can have
stable solitary wave solutions of the form
\begin{equation}
    \sol_{\Par}(x,t):= 
    \exp{\i(\frac{1}{2}v\cdot(x-a)+\gamma)}\solw(x-a),
   \label{eq:sp2a}
\end{equation}
where $\Par:=\{a,\vip,\gamma,\freq\}$, 
and $a=vt+a_0$, $\gamma=\freq t +  
\frac{v^2}{4}t+ \gamma_{0}$, with $\gamma_0 \in 
[0,2\pi)$, $a_0,\vip\in\mathbb{R}^{d}$ and
$\freq\in\mathbb{R}^{+}$, constant, 
and where $\solw$ is a positive solution of the 
nonlinear eigenvalue problem
\begin{equation}
   (-\Laplace + \freq) \solw - f(\solw) = 0  \label{eq:sp2} 
\end{equation}
(see Section~\ref{sec:res}).
Solutions of 
eq.~\eqref{eq:sp1} of the form $\eqref{eq:sp2a}$ describe solitary 
waves
traveling through space with a constant velocity $\vip$, and with an
oscillatory phase given by $\freq t-\frac{1}{4}v^2 t$. Existence of
such solutions, for a large class of nonlinearities has been
established in 
\cite{Strauss1977,Berestycki+Lions+Peletier1981,Berestycki+LionsI1983,Berestycki+LionsII1983,Berestycki+Gallouet+Kavian1983,Jones+Kupper1986,Lions1986b,Adachi2002}.  
See Section~\ref{sec:res} for an outline of results relevant for 
this paper.

In analyzing soliton-like solutions of eq.~\eqref{eq:sp1} we encounter
two length scales: the size $\propto \freq^{-1/2}$ of the support of
the function $\solw$, which is determined by our choice of initial
condition $\psi_0$,
and the length scale
$\propto (\sup |\nabla V|)^{-1}$ over which the external potential 
$V$ varies appreciably. We will assume that the ratio
\begin{equation}\label{eq:scales}
   \eps = \frac{\sup |\nabla V(x)|}{\sqrt{\freq}},
\end{equation}
is {\em small}; \ie that the potential $V$ varies little over the 
support of a solitary wave solution.

When $V$ does not vanish, the wave $\sol_{\Par}(x)$ in \eqref{eq:sp2a}  
does {\em not} solve eq.~\eqref{eq:sp1}. 
However,  we expect that if the initial condition
\begin{equation}
    \psi_0(x) := \psi(x,0) 
\end{equation}
for \eqref{eq:sp1} is close to $\sol_{\Par_0}(x)$
for some $\Par_{0}$,
in the sense that 
\begin{equation}
\nrmHo{\exp{-\i \frac{1}{2}\vip_0\cdot x}(\psi_0-\sol_{\Par_{0}})
} \leq \epsE,
\end{equation}
then for {\em all times} $0\leq t \leq \frac{T}{\eps+\epsE^2}$, where 
$T$ is some positive constant,
the solution $\psi(x,t)$ remains 
{\em close} to a solitary wave of the form \eqref{eq:sp2a}
for some {\em time-dependent} parameters $\freq,\vip,a$ and 
$\gamma$ 
determined by $V$ and $\psi_0$. 
We will show, more specifically, that 
these parameters can be chosen to be solutions of the following system
of ordinary differential equations
\begin{equation}\label{eq:new}
    \dot{a} = \vip \; , \ \ 
    \frac{\dot{\vip}}{2} = - \nabla V(a) \; , \ \ 
    \dot{\freq} = 0 \; ,
\end{equation}
and
\begin{equation}\label{eq:new2}
   \dot{\gamma} = \freq
       +\frac{\vip^2}{4}  - V(a) \; .
\end{equation}
(up to error terms of size $\Oh(\eps^2+\epsE^2)$), with initial
conditions given by $a(0)=a_0$, $\vip(0)=\vip_0$, $\freq(0)=\freq_0$,
and $\gamma(0)=\gamma_0$, with $a_0,\vip_0,\freq_0,\gamma_0$ as in
\eqref{eq:sp2a}. 
We observe that
the first two equations in \eqref{eq:new} are {\em Newton's equations
of motion} for the trajectory ($a(t),\vip(t)=\dot a(t)$) of a point
particle of mass $\frac{1}{2}$ moving in the external potential
$V(a)$.  The center of a solitary wave solution of \eqref{eq:sp1}
follows this trajectory, up to deviations $\Oh(\eps+\epsE)$ due to
``radiation damping''.

We state here rigorously the result discussed above for a special class
of nonlinearities. The general class of nonlinearities is introduced 
and discussed in the next section. 
We assume the external potential $V(x)$ satisfies the conditions   
\begin{equation}  
   V\in \C{2} \ \text{and}\ |\partial_x^\alpha V(x)| 
\leq C_\alpha \eps^{|\alpha|},
 \ \
   \text{for}\ |\alpha|\leq 2.
   \label{con:V}
\end{equation}
In \eqref{con:V}, $\eps>0$ is the small
parameter introduced in \eqref{eq:scales}. In other words, we find
it convenient to fix the size of the support (the `width') of the 
solitary wave solution at 
$\Oh(1)$ and assume that the external potential $V(x)$
varies slowly. Let $\epsilon:=\eps +\epsE$.

\begin{theorem}
    Assume that the nonlinearity $f$ is given by \eqref{eq:lnl}, 
    and assume that the potential $V$ 
    satisfies \eqref{con:V} with $\eps\ll 1$.
    Let $I_0$ be any closed, bounded interval in $(0,\infty)$. Let 
    $\epsE\ll 1$
    and the initial condition $\psi_{0}$ satisfy 
    \begin{equation}
        \nrmHo{\exp{-\i\frac{1}{2}\vip_{0}\cdot x}
        (\psi_{0}-\sol_{\Par_{0}})}<\epsE,
    \end{equation}
    for some $\Par_{0}\in 
    \mathbb{R}^{d}\times\mathbb{R}^{d}\times [0,2\pi)\times I_0$. 
    Then there is a constant $T>0$, independent of 
    $\eps$, $\epsE$ but possibly dependent on $I_0$,
    such that for times $0\leq t\leq T(\eps+\epsE^{2})^{-1}$, the 
    solution to \eqref{eq:sp1} with this initial condition is of 
    the form 
    \begin{equation}
    \psi(x,t) =\exp{\i(\frac{1}{2}\vip\cdot (x-a) + \gamma)}
        (\solw(x-a) + w(x-a,t)),
    \end{equation}
    where 
    \begin{equation}
        \nrmHo{w}=\Oh(\epsilon),
    \end{equation}
    and where the parameters $\vip, a, \gamma$ and $\freq$ satisfy the 
    differential equations
\begin{align}
    \frac{1}{2}\dot{\vip} & = - (\nabla V)(a) +
    \Oh(\epsilon^2), \\ 
    \dot{a} & = \vip + \Oh(\epsilon^2), \\
    \dot{\gamma} & = \freq - V(a) + \frac{1}{4}\vip^2 + 
    \Oh(\epsilon^2), \\ 
    \dot{\freq} &= \Oh(\epsilon^2).
\end{align}
\end{theorem}
The same conclusions hold for nonlinearities of the form 
$f(\psi)=g(|\psi|^{2})\psi+(W*|\psi|^{2})\psi$ where $W$ and $g$ 
satisfy explicit conditions
(see the discussion of the conditions in Section~\ref{sec:res}),
provided an additional spectral condition is satisfied 
(see Condition~\ref{con:g.spec} in Section~\ref{sec:res}).

The first result of this type was proved by 
Fr\"{o}hlich, Tsai and Yau \cite{Frohlich+Tsai+Yau2000,Frohlich+Tsai+Yau2002}
for the Hartree equation (\eqref{eq:sp1} with 
\eqref{eq:nlnl}) under a spectral condition 
(see Condition~\ref{con:g.spec} of Section 2).
The choice of
the Hartree type nonlinearity plays an important role in 
\cite{Frohlich+Tsai+Yau2000,Frohlich+Tsai+Yau2002}. For local, pure
power nonlinearities and a small parameter, $\eps$, 
Bronski and Jerrard 
\cite{Bronski+Jerrard2000} have shown that if an initial condition
is of form \eqref{eq:sp2a}, with $t=0$, then the solution $\psi(x,t)$ of 
\eref{eq:sp1} satisfies 
\begin{equation}
    \eps^{-d}|\psi(\frac{x}{\eps},\frac{t}{\eps})|^{2}
	\diff^{d} x 
    \rightarrow \nrmL{\solw}^{2}\delta_{a(t)}
\end{equation}
in the $\set{C}^{1*}$ (dual to $\C{1}$) topology, provided $a(t)$
satisfies the equation $\frac{1}{2}\ddot{a}=\nabla U(a)$, 
where $V(x)=U(\eps x)$ (see \eqref{eq:new}). 

Our approach is built on important developments in the nonlinear
Schr\"odinger equation (NLS) in the last 20 years (see
\cite{Cazenave1996,Sulem1999} for reviews). We outline
the landmark developments briefly here. Orbital stability of NLS
solitary waves for $V=0$ was proved by 
Cazenave and Lions \cite{Cazenave+Lions82} and
M.~Weinstein \cite{Weinstein1985,Weinstein1986}, 
whose result was significantly extended by Grillakis, Shatah and Strauss
\cite{Grillakis+Shatah+Strauss1987,Grillakis+Shatah+Strauss1990}.
The next significant step was made by A.~Soffer and
M.~Weinstein \cite{Soffer+Weinstein1990} who proved, under some
restrictive conditions, asymptotic stability of nonlinear ground
states for \eref{eq:sp1} and by V.~Buslaev and G.~Perel'man
\cite{BP92} who, motivated by \cite{Soffer+Weinstein1990}, proved
asymptotic stability of NLS solitary waves (V=0) in one dimension, again
under certain restrictive conditions. These results were significantly
extended by Tsai and Yau \cite{Tsai+Yau2002,Tsai+Yau2002b,Tsai+Yau2002c}, Soffer and Weinstein
\cite{Soffer+Weinstein1992,Soffer+Weinstein2003}, Cuccagna
\cite{Cuccagna2001,Cuccagna2002}, Buslaev and Perel'man
\cite{Buslaev+Perelman1995}, and
Buslaev and C.~Sulem \cite{Buslaev+Sulem2002}. 
Furthermore Perel'man \cite{Perelman2001}, 
and Rodnianski, Schlag and Soffer \cite{RSS}, have
obtained the first results on soliton scattering.  
Many of the issues touched upon in the present paper were studied
also in work of Gustafson and Sigal \cite{GS} on
dynamics of magnetic vortices.

We also mention 
interesting non-rigorous results by Pelinovsky, Afanasjev and  
Kiv\-shar \cite{Pelinovsky1996}
and Pelinovsky and Grimshaw \cite{Pelinovsky+Grimshaw1997} on dynamics 
of NLS solitons near the borderline for stability.
Nonrigorous results on dynamics of `center of 
mass' of solitons were obtained by E.~van Groesen and F.~Mainardi 
and G.~Derks and E.~van Groesen 
\cite{Groesen+Mainardi,Derks+Groesen}.
  
Our paper is organized as follows.
In Section~\ref{sec:res} we present the general hypotheses on the class of 
nonlinearities and formulate our main result under these hypothesis.
In Section~\ref{sec:sp}, we explain the Hamiltonian and variational 
aspects of the dynamics given by eq.~\eqref{eq:sp1} which play a role
in our proof.  
In Section~\ref{sec:man}, we describe the symmetries of 
\eref{eq:sp1} for $V\equiv 0$ 
and find the ``zero modes'' associated with these 
symmetries. Furthermore, we introduce and analyze a 
finite-dimensional manifold, $\Mf$, of stable solitary wave 
solutions to \eref{eq:sp1}.
In Section~\ref{sec:on}, we introduce a convenient parameterization of 
functions in a small neighborhood of $\Mf$ in phase space.
In Section~\ref{sec:mf}, we transform the equation 
\eqref{eq:sp1} to a moving frame, and then rewrite the resulting 
equation in terms of the parameters introduced in Section~\ref{sec:on}
In Sections~\ref{sec:Lf} and~\ref{sec:lL}, we control solutions of our
equations of motion in a moving frame by constructing an approximately
conserved Lyapunov functional. 
The proof of our main result is completed in Section~\ref{sec:MT}. 
Some material of technical or review nature is collected in four
appendices.

Remark about the notation: in this paper we consider equations, maps and 
functionals on complex spaces such as $\Hone(\mathbb{R}^{d},\mathbb{C})$, 
which sometimes are identified with real spaces; \eg 
$\Hone(\mathbb{R}^{d},\mathbb{R})=
\Hone(\mathbb{R}^{d},\mathbb{R})\oplus\Hone(\mathbb{R}^{d},\mathbb{R})$,
under the identification $\psi \leftrightarrow (\RE \psi,\IM \psi)$. 
In this case the operator of multiplication by $\i^{-1}$ is 
identified with the operator 
\begin{equation}
    \J = \begin{pmatrix} 0 & 1 \\ -1 & 0 \end{pmatrix}
\end{equation}
(which defines a complex structure on the corresponding real spaces).
Thus a real function $\solw$ is sometimes written as $(\solw,0)$ and 
similarly an imaginary function as $\i \solw$ as $(0,\solw)$ (or even 
as $\J\solw$).  Fr\'echet derivatives are always understood to
be defined on real spaces. They are denoted by primes. $c$ will denote 
various constants which may depend only on the interval $I_0$ for 
the parameter $\freq_0$.

\section*{Acknowledgments}

B.L.G.J. and I.M.S. 
are grateful to J. Colliander for useful discussions and
remarks and to ETH-Z{\"u}rich for hospitality during their work on 
this paper. J.F. thanks T.-P. Tsai and H.-T. Yau for very useful
discussions and correspondence which led to the results in
\cite{Frohlich+Tsai+Yau2000,Frohlich+Tsai+Yau2002}.

\section{Main Result}
\label{sec:res}

In this section we formulate general assumptions on the nonlinearity $f$ in 
\eqref{eq:sp1} and state our main result under general
assumptions. Our assumptions are rather abstract, each responsible for
certain aspect of the problem and of our approach. Then we discuss 
specific nonlinearities for which these assumptions are satisfied
\renewcommand{\theenumi}{(\Alph{enumi})}
\begin{enumerate}
   \item \label{con:g.ham} (Energy) There exists a 
   $\C{3}$-functional $F:\Hone\mapsto \mathbb{R}$ such that 
   $F'(\psi)=f(\psi)$, $\sup_{\nrmHo{u}\leq 
   M} \nrm{F''(u)}_{\mathrm{B}(\Hone,\Sob{-1})}<\infty$ and 
   $\sup_{\nrmHo{u}\leq M}
   \nrm{F'''(u)}_{\Hone\mapsto \mathrm{B}(\Hone,\Sob{-1})}\leq 
   \infty$ (here $\mathrm{B}$ is the space of bounded linear operators);

   \item 
   \label{con:g.sym}
   (Symmetries) 
   $F(\mathcal{T}\psi)=F(\psi)$, where $\mathcal{T}$
   is a translation $\Tr:u(x)\mapsto u(x-a)$ $\forall a\in 
   \mathbb{R}^{d}$, a rotation $\SRot:u(x)\mapsto 
   u(R^{-1}x)$ $\forall R\in \RotSpace(d)$, a gauge transform 
   $\Rot:u(x)\mapsto \exp{\i \gamma}u(x)$ $\forall \gamma\in[0,2\pi)$,
   a boost transform $\Ph:u(x)\mapsto \exp{\frac{\i}{2}v\cdot 
   x}u(x)$ $\forall v\in\mathbb{R}^{d}$, 
   or a complex conjugation $\CC: u(x)\mapsto \bar{u}(x)$;

   \item \label{con:g.Omega} (Existence of solitons)
      There is an interval $I\subset \mathbb{R}$ 
        such that $\forall \freq\in I$
          \eref{eq:sp2} has a positive, spherically symmetric,
        $\Ltwo\cap \C{2}$ solution $\solw$, such that 
        $\nrm{|x|^{3}\solw}+\nrm{|x|^{2}|\nabla\solw|}+
        \nrm{|x|^{2}\partial_{\freq}\solw}<\infty$ $\forall \freq\in I$
        (here $\nrm{\cdot}$ denotes the $L^2$ norm);
   \item \label{con:g.stab} (Orbital stability) The
   solutions $\solw$, $\freq\in I$, of \eqref{eq:sp2}  
   described in~\ref{con:g.Omega}, satisfy
          \begin{equation}
              \partial_\freq \int \solw^{2} \diff^d x>0;   \label{eq:Qwbbl}
          \end{equation}
          
   \item \label{con:g.sol} (GWP)
     \Eref{eq:sp1} is globally well-posed in $\Sob{1}$ and in $\Sob{2}$;

   \item \label{con:g.spec} (Null space condition) $\forall \mu\in I$, 
   \begin{equation}
        \Null{\LLm} = \Span\{ \big(0, \solw\big), \big(\partial_{x_j}\solw, 
        0\big), j=1,...,d \}\; 
       \label{eq:NullL}
   \end{equation}
   where $\LLm:= -\Laplace +\freq -f'(\solw)$,
   the Fr\'echet derivative of the map $\psi\mapsto (-\Laplace + 
   \freq)\psi-f(\psi)$ at $\solw$.
\end{enumerate}
\renewcommand{\theenumi}{\roman{enumi}}
We now discuss conditions~\ref{con:g.ham}--\ref{con:g.spec}, 
beginning with general remarks.
Condition~\ref{con:g.ham} 
allows us to define the conserved energy or Hamiltonian
\begin{equation}\label{eq:HV}
    \Ham_{V}(\psi) = \frac{1}{2} \int (|\nabla \psi|^{2} + 
    V|\psi|^{2}) \diff^{d} x - F(\psi),
\end{equation}
and  gives us the following estimates on the nonlinearities
\begin{equation}\label{R}
    |\RII{w}|\leq 
    c(M)\nrmHo{w}^{2}, \ \  |\RIII{w}|\leq c(M)\nrmHo{w}^{3},
\end{equation}
and 
\begin{equation}\label{eq:N}
    \nrmD{\NII{w}}\leq C(M)\nrmHo{w}^{2}
\end{equation}
for any $\sol\in\Htwo(\mathbb{R}^{d})$ 
and $w\in \Hone(\mathbb{R}^{d})$.
with $\nrmHo{\sol}+\nrmHo{w} \leq M$.
Here 
\begin{eqnarray}
  \RII{w}:=F(\sol+w)-F(\sol)-\dotp{F'(\sol)}{w}, \\
  \RIII{w}:=F(\sol+w)-F(\sol)-\dotp{F'(\sol)}{w}-
  \frac{1}{2}\dotp{F''(\sol)w}{w} \label{eq:RIII}
\end{eqnarray}
and 
\begin{equation}
\NII{w}=F'(\sol+w)-F'(\sol)-F''(\sol)w.
\end{equation}
Note that 
\begin{equation} 
    \NII{w}=\pRIII{w}.
\end{equation}
We can assume without loss 
of generality that $f'(0)=0$. Then one can show that $I\subset \mathbb{R}_+$. 

As was
mentioned in the introduction, the nonlinearities of interest to us
are local nonlinearities, 
\begin{equation}\label{eq:l2}
f(\psi)(x)=h(|\psi(x)|^2)\psi(x),
\end{equation} 
for some real function $h$
on $\mathbb{R}_{+}$, and Hartree-type nonlinearities, 
\begin{equation}\label{eq:nl2}
f(\psi) = (W * 
|\psi|^{2})\psi,
\end{equation}
where $W$ is a fixed, real valued, spherically symmetric 
function, tending to 0 at $\infty$. 
More generally we consider nonlinearities the form
\begin{equation}
   f(\psi)(x) = h(|\psi(x)|^2)\psi(x) + (W*|\psi|^2)(x)\psi(x). \label{eq:non}
\end{equation}
We discuss now under which conditions on $h$ and $W$ in \eqref{eq:non} 
Conditions~\ref{con:g.ham}--\ref{con:g.spec} are satisfied.

Condition~\ref{con:g.ham}. For nonlinearities of type \eqref{eq:non} the 
functional $F$ is given by 
\begin{equation}
    F(\psi) = \frac{1}{2} \int H(|\psi|^{2}) 
    +\frac{1}{2}(W*|\psi|^{2}) \diff^d x,
\end{equation}
where $H(s)=\int_{0}^{s}h(p)\diff p$. The functional $F$ is $\C{3}$ with the 
inequalities specified in Condition~\ref{con:g.ham} satisfied,
provided $h(s)$ is $\C{2}$ with 
$h^{(k)}(s) \leq c(1+s^{q-k})$ ($k=0,1,2$), 
for some $q < 2/(d-2)$, and $d<4$, 
and $W \in \Lp{p} + \Lp{\infty}$ for some $p > d/2$.

Condition~\ref{con:g.sym}. 
This condition is trivially satisfied for \eqref{eq:non}
with $W(x)=W(|x|)$.

Condition~\ref{con:g.Omega}. Existence of a positive, spherically symmetric
solitary wave solution to Eq \eqref{eq:sp2}, was proved in 
\cite{Berestycki+LionsI1983,Berestycki+LionsII1983,Berestycki+Lions+Peletier1981} 
for local nonlinearities \eqref{eq:l2} satisfying
\begin{align}
   -\infty < \lim_{r\rightarrow 0} h(r) < \freq, \nonumber \\
   -\infty \leq\lim_{r\rightarrow\infty}r^{-\alpha}h(r)
   \leq C ,  \label{eq:Uclass1}
\end{align}   
for $0<\alpha < 2/(d-2)$, if $d>2$ and
$\alpha \in(0,\infty)$, if $d=1,2$,   
\begin{equation*}
\exists \zeta>0, \ \text{such that}\ 
        \int^\zeta_0  h(r) \diff r>\freq\zeta.  \nonumber
\end{equation*}
Additional results, for a large class of nonlinearities can be found in 
\cite{Strauss1977,Berestycki+Gallouet+Kavian1983,Jones+Kupper1986,Lions1986b,Adachi2002}. 

For nonlocal nonlinearities \eqref{eq:nl2} 
(Hartree equation) with 
\begin{equation}
    W\in \Lp{p}_{\mathrm{loc}},\ p\geq\frac{d}{2},\ 
W\rightarrow 0\ \text{as}\ |x|\rightarrow \infty, \label{eq:Uclass2}
\end{equation}
existence of solutions was proved in 
\cite{Cazenave1996,Ginibre+Velo1980,Frohlich+Tsai+Yau2002,Li1990,Li+Li1993,Adachi2002}.

Moreover, for nonlinearities \eqref{eq:l2} or \eqref{eq:nl2}
satisfying conditions \eqref{eq:Uclass1} or 
\eqref{eq:Uclass2}, solutions to \eqref{eq:sp2} (solitary waves) are 
exponentially decaying at $\infty$ as $\Oh(\exp{-\sqrt{\freq}|x|})$;
see \cite{Berestycki+LionsI1983,Peletier+Serrin1983}.

Condition~\ref{con:g.stab}. Condition \eqref{eq:Qwbbl} is a sufficient
condition for orbital stability of solitary waves for (generalized)
nonlinear Schr\"{o}dinger equations (see
\cite{Grillakis+Shatah+Strauss1987,Grillakis+Shatah+Strauss1990}).
This condition is to be checked for each nonlinearity.  In the special
case of a pure power nonlinearity,  
$f(\psi)=\lambda|\psi|^{2s}\psi$, we
have $\solw(x) = \freq^{\frac{1}{2s}}\sol_{\freq=1}(x\sqrt{\freq})$.
Thus condition~\ref{con:g.stab} reduces to the condition that $s<2/d$.

Condition~\ref{con:g.sol}.
Assume that $f$ is of the form \eqref{eq:non} with 
$h: [0,\infty)\mapsto \mathbb{R}$, smooth and 
satisfying $h(0)=0$, and 
\begin{equation}
    |h'(r)|\leq C (1+r^{\alpha-1}) \; , \label{eq:hp}
\end{equation}
for some $\alpha\in [0, \frac{2}{d-2})$, 
for $d\geq 3$ and $\alpha\in [0,\infty)$ for $d=1,2$ and
\begin{equation}
   h(r)\leq C(1+r^{\beta}) \; ,   \label{eq:hpp}
\end{equation}
for some $\beta\in [0,\frac{2}{d})$. Furthermore assume that 
$W:\mathbb{R}^d\mapsto \mathbb{R}$
is an even potential such that 
\begin{equation}
   W\in \Lp{q}(\mathbb{R}^d) + \Lp{\infty}(\mathbb{R}^d),
\end{equation} 
for some $q\geq 1$, $q>d/4$ and let $W_+=\max(0,W)$ s.t.
\begin{equation}
   W_+ \in \Lp{r}(\mathbb{R}^d) + \Lp{\infty}(\mathbb{R}^d),
\end{equation}
for some $r\geq 1$, $r\geq d/2$ for $d>2$ and $r>1$ for $d=2$. 
Then \eqref{eq:sp1} is globally well-posed in $\Hone$ and $\Htwo$
(see \cite[Chapter 6]{Cazenave1996}).

Condition~\ref{con:g.spec}. This condition is more delicate.
First we observe that 
\begin{equation}
  (\partial_{x_j} \solw,0), (0,\solw) 
\in \Null{\LLm}\; , \ \ \forall j=1,...d \; ,
  \label{eq:nulls}
\end{equation}
due to the fact that $\solw$ breaks the translation and gauge symmetry of
\eqref{eq:sp2} (see Section~\ref{sec:man}). 

Now we list some facts which are discussed in Appendix~\ref{app:nullL}.
If $\solw$ is spherically symmetric and $f$ is a local nonlinearity,
$f(\psi)(x)=f(\psi(x))$, then $\LL$ can have at most one 
zero eigenfunction in addition to \eqref{eq:nulls}. 
Hence 
\begin{equation} 
    d+1\leq \dim \Null{\LLm} \leq d+2.
\end{equation}    
This extra zero 
eigenfunction is spherically symmetric and is also a zero eigenfunction of the
ordinary differential operator
\begin{equation}
   A_{\freq,0} = - \Laplace_r +\freq - f'(\solw)
\end{equation}
on $\Ltwo(\mathbb{R}_+, r^{d-1}\diff r)$, where $\Laplace_r$ is the 
radial Laplacian,  
\begin{equation}
  \Laplace_r = \partial_r^2 + \frac{d-1}{r}\partial_r \; .
\end{equation}
Thus
\begin{equation}
   \Null{\LLm} = \Span\{(0, \solw),
   (\partial_{x_j}\solw, 0),(\Null{A_{\freq,0}},0) \}. 
   \label{eq:nullL}
\end{equation}

For local nonlinearities \eqref{eq:l2}, $\Null{A_{\freq,0}}=\{0\}$, 
if either $d=1$ or 
\begin{equation}\label{Uclass}
    h'(r)+h''(r)r^2>0.
\end{equation}
Alternative conditions on $h$ for $d>1$ are given in Appendix~\ref{app:nullL}.
Thus, for local nonlinearities, if either $d=1$ or 
\eqref{Uclass} holds, then condition 
\ref{con:g.spec} is satisfied (see also
\cite{McLeod+Serrin1987,Weinstein1985,McLeod1993,Stuart2001}).

In any case this extra degeneracy, if it happens,
is non-generic. If \eqref{eq:NullL} holds for some $f$, then it also holds for
small perturbations of $f$. On the other hand we expect that 
if \eqref{eq:NullL} fails 
for a given $f$ then
there are arbitrarily small perturbations of $f$ 
for which \eqref{eq:NullL} is satisfied.

It is easy to check that Condition~\ref{con:g.ham}--\ref{con:g.spec} 
are satisfied for nonlinearity \eqref{eq:lnl}.

Let $\setPar:=\mathbb{R}^d\times \mathbb{R}^d\times 
[0,2\pi)\times I$ and $\setPar_0:=\mathbb{R}^d\times \mathbb{R}^d\times 
[0,2\pi)\times I_0$, 
where $I$ is a closed, bounded interval on the positive real axis, 
specified in Condition~\ref{con:g.Omega}, and 
$\bar{I_0}$ is a closed interval contained in $I\backslash\partial I$.
Recall $\epsilon:=\eps+\epsE$.
Our main result is 
\begin{theorem}\label{thm:main}
Assume \eqref{con:V} and
\ref{con:g.ham}--\ref{con:g.spec} are satisfied. Given $\epsE>0$ and $\eps>0$ 
defined in \eqref{eq:scales} such that $\epsilon\ll 1$
and an initial condition $\psi_0$ satisfying
\begin{equation}
    \nrmHo{\exp{-\frac{\i}{2}\vip_0 \cdot x}(\psi_0 - \sol_{\Par_{0}})} 
        \leq \epsE  \; ,
\end{equation}
for some $\Par_0:=\{a_0, \vip_0, \gamma_0, \freq_0\} \in \setPar_0$.  
Then there is a constant $T>0$ (independent of
$\eps, \epsE$ but possibly dependent on $I_0$) 
such that for times $t(\eps+\epsE^2) \leq T$ the
solution to \eqref{eq:sp1} with this initial condition can be
written in the form
\begin{equation}
    \psi(x,t) =\exp{\i(\frac{1}{2}\vip\cdot (x-a) + \gamma)}
        (\solw(x-a) + w(x-a,t))
\end{equation}
where 
\begin{equation}
    \nrmHo{w} = \Oh(\epsilon)
\end{equation}
and the parameters $\vip, a, \gamma$ and $\freq$ satisfy the differential
equations
\begin{align}
    \frac{1}{2}\dot{\vip} & = - (\nabla V)(a) +
    \Oh(\epsilon^2), \\ 
    \dot{a} & = \vip + \Oh(\epsilon^2), \\
    \dot{\gamma} & = \freq - V(a) + \frac{1}{4}\vip^2 + 
    \Oh(\epsilon^2), \\ 
    \dot{\freq} &= \Oh(\epsilon^2).
\end{align}
\end{theorem}
The proof of this theorem is given in Section~\ref{sec:MT}.

\section{The Hamiltonian and variational structure of 
generalized nonlinear Schr\"odinger equations}
\label{sec:sp}

It is well known that the generalized nonlinear Schr\"odinger 
equation is a Hamiltonian equation of motion on an 
infinite-dimensional phase space. In this paper we make extensive 
use of this fact and of the symplectic geometry of phase space, 
in particular of certain ``submanifolds'' in this space. 
We outline briefly some facts which are relevant for us.

In this section we consider the generalized nonlinear Schr\"odinger 
equation \eref{eq:sp1},
\begin{equation*}
   \i \partial_t \psi = (-\Laplace + V)\psi - f(\psi) \; ,
\end{equation*}
under Conditions~\ref{con:g.ham} and~\ref{con:g.sym} only.

We study \eref{eq:sp1} on the space $\Hone(\mathbb{R}^{d},\mathbb{C})$. 
This space, considered as a real space 
($\Hone(\mathbb{R}^{d},\mathbb{R}^{2})=\Hone(\mathbb{R}^{d},\mathbb{R})
\oplus\Hone(\mathbb{R}^{d},\mathbb{R})$, 
$\psi \leftrightarrow (\RE \psi,\IM \psi)$), 
and equipped with the symplectic form
\begin{equation}
    \tfrm{u}{v} := \IM \int u\bar{v} \diff^{d}x
\end{equation}
(defined for $u$, $v$ in the tangent space at a given point 
in $\Hone(\mathbb{R}^d;\mathbb{R}^2)$,
which we can identify with
$\Hone(\mathbb{R}^{d},\mathbb{R}^{2})$ in our case),
is a symplectic space.

Define the Hamiltonian functional on 
$\Hone(\mathbb{R}^{d},\mathbb{C})$ as
\begin{equation}\label{eq:haM}
    \Ham_{V}(\psi):=\frac{1}{2} \int |\nabla \psi|^{2} + V|\psi|^{2} 
    \diff^{d} x -F(\psi),
\end{equation}
where $F(\psi)$ is as in Condition~\ref{con:g.ham}; \ie 
$F'(\psi)=f(\psi)$.
With these definitions 
\eref{eq:sp1} can be written as 
\begin{equation}
   \partial_{t}\psi = J \Ham_{V}'(\psi),
\end{equation}
where $J$ is the operator on 
$\Hone(\mathbb{R}^{d},\mathbb{R}^{2})$ (strictly speaking 
$J:(\set{T}_{\psi}\Hone)^{*}\mapsto\set{T}_{\psi}\Hone$) 
given by
\begin{equation}\label{eq:Jreal}
\J:=\begin{pmatrix} 0 & 1 \\ -1 & 0 \end{pmatrix}
\end{equation}    
in block-diagonal notation.
Thus $\J$ is a complex structure on $\Hone(\mathbb{R}^{d},\mathbb{R}^{2})$
corresponding to the operator $\i^{-1}$ on $\Hone(\mathbb{R}^{d},\mathbb{C})$.

Observe that the space $\Hone(\mathbb{R}^{d},\mathbb{C})$ also has 
a real inner product (Riemannian metric) 
\begin{equation}
    \dotp{u}{v}:=\RE \int u\bar{v}\diff^{d}x,
\end{equation}
so that $\tfrm{u}{v}=\dotp{u}{\J^{-1}v}$.

Since the Hamiltonian $\Ham_{V}(\psi)$ is autonomous (it does not 
explicitly depend on time $t$) and invariant under the gauge 
transformation $\Ham_{V}(\exp{\i \gamma}\psi)=\Ham_{V}(\psi)$ for 
all $\gamma\in [0,2\pi)$, we have 
conservation of energy, $\Ham_{V}(\psi)=\text{const}$, and ``mass'' or 
``number of particles'', $\Nn(\psi)=\text{const}$, where
\begin{equation}
    \Nn(\psi) := \frac{1}{2}\int |\psi|^{2} \diff^{d} x
\end{equation}
under the evolution of \eref{eq:sp1}. 

Due to Condition~\ref{con:g.stab} the
solitary wave profiles $\solw$
described in Condition~\ref{con:g.Omega} are local minimizers 
of the Hamiltonian $\Ham_{V=0}(\psi)$
restricted to the spheres 
\begin{equation}
    \{\psi\in\Hone: N(\psi)=\mm\}
\end{equation}
for $m>0$ (see \cite{Grillakis+Shatah+Strauss1987}, Thm 3).
Hence they are critical points of the functional 
\begin{equation}\label{eq:EW}
    \Ew(\psi):= \frac{1}{2}\int |\nabla \psi|^2 + \freq|\psi|^2 \diff^d x- 
   F(\psi),
\end{equation}
where $\freq = \freq(m)$ 
is a Lagrange multiplier, and \eqref{eq:sp2} is just the
Euler-Lagrange equation for $\Ew(\psi)$. 

Observe that the functional 
$\Ew(\psi)$ also arises as 
$(t_{1}-t_{0})\Ew(\phi)=\Sn_{V=0}(\phi \exp{\i \freq t})$
for any $\phi(x)$, where $\Sn_{V}$ is the action for \eref{eq:sp1}:
\begin{equation}
   \Sn_{V}(\psi) :=\int_{t_{0}}^{t_{1}} \left ( 
   \frac{1}{2}\int_{\mathbb{R}^{d}} \IM \dot{\psi}\bar{\psi}\diff^{d} 
   x+\Ham_{V}(\psi) \right) \diff t.
\end{equation}

The functional $\Ew(\psi)$ will play an important role in our approach. 
We will use it as a Lyapunov functional in estimating the 
fluctuations $w$. Finally, we note that the operator $\LL$ that appears in 
Condition~\ref{con:g.spec} is the Hessian of $\Ew(\psi)$ at $\solw$: 
$\LL:=\Ew''(\solw)$.

\section{Symmetries, zero modes, and the manifold of solitary waves}
\label{sec:man}

In this section we introduce the manifold of solitary waves which is
obtained by applying the generalized symmetry transforms (see below)
to a fixed solitary wave. The tangent space to the manifold is
introduced, and its inherited symplectic
form is derived. Furthermore we prove the key fact that the
inherited symplectic form is non-degenerate.  

Starting from this section we will often use the abbreviation 
$\sol\equiv \solw$.

\Eref{eq:sp1} with $V\equiv 0$ is invariant
under spatial translations $\Tr$, gauge transformations $\Rot$, and
Galilean transformations $\Bo$, where
\begin{align}
    \Tr &:\psi(x,t) \mapsto \psi(x-a,t) \; , && 
    \Rot : \psi(x,t)\mapsto \exp{\i\gamma}\psi(x,t) 
        \label{eq:T1} \; , \\ 
    \Bo &: \psi(x,t)\mapsto 
    \exp{\i(\frac{1}{2}\vip\cdot x - \frac{1}{4}|\vip|^2 t) }
    \psi(x-\vip t,t)\; .
        \label{eq:T2}
\end{align}
(Transformations \eqref{eq:T1}--\eqref{eq:T2} 
map solutions of eq.~\eqref{eq:sp1} with $V=0$ 
into solutions of \eqref{eq:sp1} with $V=0$.)
The symmetries $\Tr$, $\Rot$, $\Bo$, have associated
conserved quantities: field momentum, mass and 'center of mass motion', 
(\cf \cite{Frohlich+Tsai+Yau2002}),
\begin{equation}
\int \i \bar{\psi}\nabla\psi \diff^d x, \ \ \int |\psi|^2 \diff^d x, \ \ 
\int \bar{\psi} (x+2\i t \nabla)\psi \diff^d x.
\end{equation}

For $f(\psi)=|\psi|^{2\spow}\psi$ 
and $V=0$, \eref{eq:sp1} is also invariant under 
the scaling transformation 
\begin{align}
    \Sc : \psi(x,t) \mapsto \freq^{\frac{1}{2\spow}} 
                \psi(\sqrt{\freq} x,\freq t)\; .
\end{align}

When the external potential is introduced into the problem, the
translational and Galilean invariance are broken. In particular,
conservation of the field momentum 
is replaced by the following `Newton's law' 
(Ehrenfest's theorem)
\begin{equation}\label{eq:Ehrenfest}
\partial_t \dotp{\psi}{-\i \nabla \psi} = - \dotp{\psi}{(\nabla V)\psi},
\end{equation}
which plays an important role in our
analysis, and which is proved in Appendix~\ref{app:Ehrenfest}.

Let $\Ph:\psi(x)\mapsto \exp{\frac{\i}{2}\vip\cdot x}\psi(x)$, be the
boost transform.  We introduce the combined symmetry transformations
$\TB$:
\begin{equation}
   p(x)\mapsto p_{a\vip \gamma} :=\TB p = 
        \Tr \Ph\Rot p(x) =
        \exp{\i (\frac{1}{2}\vip \cdot (x-a)+\gamma)}p(x-a) \; .
\end{equation}

Let $\sol_{a\vip\freq\gamma}:=\TB \solw$. 
We define the manifold of solitary waves as
\begin{equation}
   \Mf := \{\sol_{a\vip \gamma\freq}
                : a,\vip, \gamma,\freq \in 
   \mathbb{R}^d\times 
   \mathbb{R}^d\times [0,2\pi]\times I \} \; .
\end{equation}
The tangent space to this manifold at the solitary wave 
profile $\solw\in \Mf$ is given by
\begin{equation}
   \set{T}_{\solw}\Mf = \Span(\za,\zg,\zt,\zn) \; ,
\end{equation}
where
\begin{align}
   \tr :=&\left.\nabla_a  \Tr \solw \right|_{a=0} = -\nabla \solw\; ,
   \label{eq:t1} \\
   \gu :=& \left. \frac{\partial}{\partial \gamma} 
        \Rot \solw\right|_{\gamma=0} = \i \solw\; ,\\
    \ga := &\left. 2 \nabla_\vip \Bo \solw \right|_{\vip=0,t=0} = 
        \i x\solw \; ,
\end{align}
and
\begin{equation}\label{eq:t4}
    \sa := \partial_{\freq} \solw \; .
\end{equation}

Symmetries of \eqref{eq:sp1} which are broken by the 
solitary wave solution $\solw\exp{\i \freq t}$ lead to zero modes and 
``symplectically associated zero modes'' of the Hessian $\LL$.
Namely we have for $\sol=\solw$
\begin{equation}
   \LL\tr = 0 \; , \ \mbox{and} \ \LL\gu = 0 \; , \label{eq:trgu}
\end{equation}
and
\begin{equation}
        \LL\ga = 2 \i \tr \, , \ \ \mbox{and} \ \LL\sa = \i  \gu \; .
    \label{eq:gasa}
\end{equation}
The functions $\ga$ and $\sa$ are zero modes 
for the operator $(\i \LL)^2$, \ie
\begin{equation}
        \i \LL(\i \LL\ga) = 0 \; , \ \ \mbox{and} \ \i \LL(\i \LL\sa)= 
        0 \; .
\end{equation}

The relations \eqref{eq:trgu} are proved 
by taking the derivatives of the equation $\Ew'(\Tr\Rot\solw)$ $=0$ 
with respect to the parameters $a$ and $\gamma$, at $a=0$ and $\gamma=0$,
and similarly for \eqref{eq:gasa}.

We have shown above that 
\begin{equation}
        \TM \subset \gN(\J \LL ) \; ,
\end{equation}
where 
\begin{equation}
        \gN(\J \LL) := \Span\{ \bigcup_{n\geq 1} \Null{(\J \LL)^n}\} \; .
\end{equation}

In what follows we denote 
$\Par:=\{a,\vip,\gamma,\freq\}$ and $\eta_\Par:=\eta_{a\vip\gamma\freq}$.
The manifold $\Mf$ inherits a symplectic form from $(\Hone,\omega)$.
This symplectic form is determined by the operator
\begin{equation}\label{eq:iom}
    \piom:=\J^{-1}\upharpoonright_{\set{T}_{\sol_{\Par}}\Mf}\equiv\Pro 
    J^{-1}\Pro\upharpoonright_{\set{T}_{\sol_{\Par}}\Mf},
\end{equation}    
where $\Pro:\set{T}_{\sol_{\Par}}\Hone\mapsto \set{T}_{\sol_{\Par}}\Mf$ is the 
$\Ltwo$-orthogonal projection onto $\set{T}_{\sol_{\Par}}\Mf$. Namely 
$\rfrm{u}{v}:=\dotp{u}{\piom v}$. The key fact here is that this
symplectic form is non-degenerate, \ie the operator 
$\J^{-1}\upharpoonright_{\set{T}_{\sol_{\Par}}\Mf}$, 
is invertible $\forall \sol_{\Par}\in\Mf$, as 
shown in Lemma~\ref{lem:Oinv} below. 
Define
\begin{equation}\label{eq:mw}
\mw:=\frac{1}{2}\int |\solw|^2 \diff^d x.
\end{equation}

\begin{lemma}\label{lem:Oinv}
If $m'(\freq)>0$, then $\piom$ is invertible. 
\end{lemma}
Note that $m'(\freq) > 0$ is 
exactly the assumption for stability in
Section~\ref{sec:res}, condition~\ref{con:g.stab}. 
\begin{proof}
We prove that $\iom$ is invertible by showing that its matrix 
$\Miom:=(\dotp{z_{j}}{\J^{-1}z_{k}})$ in the basis
$\{z_{1},\ldots,z_{2d+2}\} \equiv \{\tr,\ga,\gu,\sa\}$ in $\TM$ is 
invertible. We compute 
\begin{equation}\label{eq:eiom}
   \left.\iom\right|_{\{z_{k}\}} = \begin{pmatrix} 
      0 & -\mm\bone & 0 & 0 \\ 
      \mm\bone & 0 & 0 & 0 \\
      0 & 0 & 0 & \mmp \\
      0 & 0 & -\mmp & 0
 \end{pmatrix}, 
\end{equation}
This matrix is invertible provided $\mmp\neq 0$ 
($m>0$ always).
Since $\Omega^{-1}_{\sol_\Par}$ is related to $\iom$ by a similarity
transform (see \eqref{eq:dk}), it is invertible as well. \qed
\end{proof}

\begin{corollary}\label{cor:nondeg}
$\forall z\in\TM$, there exists at least one element, $z'\in \TM$,
such that $\stfrm_\sol(z,z')\neq 0$.
\end{corollary}
\begin{proof}
This follows directly from the non-degeneracy of $\iom$.\qed
\end{proof}

\begin{remark}
The real form of the vectors \eqref{eq:t1}--\eqref{eq:t4} is 
\begin{equation}
(-\nabla \sol , 0 ), \ \ (0 , \sol ), \ \ 
(  0 , x \sol ), 
\ \ (\partial_\freq \sol , 0 ).
\end{equation}
We abuse notation and denote both the real 
and complex representations of the 
vectors as $-\nabla \sol$, $-\J \sol$, $-\J x \sol$ and
$\partial_\freq \sol$, where 
for the real representation we interpret $\sol$ as $(\sol,0)$, and 
$\J$ is of the form \eqref{eq:Jreal}, whereas in the complex notation
$\J=\i^{-1}$.
\end{remark}
\begin{remark}
  For the special case with $f(\psi)=\lambda |\psi|^{2\spow}\psi$, 
\begin{equation}
     \sa :=\left.\partial_{\freq} 
         \Sc\sol \right|_{\freq=1,t=0}= 
        \frac{1}{2}(\frac{1}{\spow} + x\cdot \nabla )\sol = 
        \partial_\freq \sol_{\freq} \; .
\end{equation}
\end{remark}
\begin{remark}
Equation \eqref{eq:sp1} with $V=0$ and with a nonlinearity of the
form \eqref{eq:non}, where $W$ is spherically symmetric has also
rotational symmetry: $\psi(x,t)\mapsto \psi(Rx,t)$, $R\in
\RotSpace(d)$. This symmetry does not play a role in our analysis since 
we consider only spherically symmetric solitary wave solutions of 
\eqref{eq:sp2}.
\end{remark}
\begin{remark}
Note that for $f(\psi)=\lambda |\psi|^{2\spow}\psi$ we can 
put the $\freq$ dependence on the 
same footing as the transformation induced for 
$\vip$, $a$ and $\gamma$. 
Indeed, define
\begin{equation}
    \Tb = \TB \circ \Sc.
\end{equation}
Then $\sol_{a\vip\gamma\freq}=\Tb \sol_1$.
\end{remark}

\section{Skew orthogonal decomposition}
\label{sec:on}

In this section we introduce the skew orthogonal decomposition
of a solution $\psi$ along the manifold $\Mf$ and
in the skew orthogonal direction (see \cite{Arnold1989}), 
and show that this uniquely defines the
solitary wave solution parameters appearing in the decomposition.
 
We recall the notation $\Par = \{\gamma, a, \vip, \freq\}$ so that
$\solp := \sol_{a \vip \gamma\freq}$.   Define the $\delta$-neighbor\-hood 
\begin{equation}
     U_{\delta} = \{\psi\in \Hone:\inf_{\Par\in \setPar_0}
       \nrmHo{\psi-\sol_\Par} \leq \delta\} \; 
\end{equation}
of the manifold $\Mf':=\{\solp:\Par \in \setPar_0\}$, where
\begin{equation}
   \setPar_0:= \mathbb{R}^d\times\mathbb{R}^d\times [0,2\pi)\times I_0.
\end{equation}
Here $I_0$ is any bounded 
closed interval contained in $I\backslash \partial I$, (for the definition
of $I$ Condition~\ref{con:g.Omega}).

The main result of this section is 
\begin{proposition}\label{prop:IFT}
Let $\psi \in U_{\delta}$. For $\delta>0$ sufficiently small,  
there exists a unique 
$\Par = \Par(\psi)\in \C{1}(U_{\delta},\setPar)$ 
such that 
\begin{equation}
        \tfrm{\psi-\sol_{\Par}}{z}=0 \ \text{\ie}\
        \rdot{\psi-\sol_{\Par}}{\J^{-1} z} = 0 \; , \forall z \in
        \mathrm{T}_{\sol_{\Par}}\Mf \; .
        \label{eq:od3}
\end{equation}
\end{proposition}

\begin{proof}
We will use the following notation:
\begin{eqnarray}
        \Par:=\{\Par_1,...,\Par_{2d+2}\}:=\{a,\vip,\gamma,\freq\},\\
   \{z_{\freq,1},...,z_{\freq,2d+2}\}:= \{\tr, \ga, \gu, \sa\},
\end{eqnarray}
and 
\begin{equation}\label{eq:dk}
   \partial_k = \partial_{\Par_k},\ k=d+1,...,2d+2,\
   \partial_k = \partial_{\Par_{k}}+\frac{1}{2}\Par_{k+d} 
   \partial_{\Par_{2d+1}},\ k=1,...,d.
\end{equation}
Denote $z_{\Par,j}:=\TB z_{\freq,j}$ so that 
$\partial_k \solp = z_{\Par,k}=\TB z_{\freq,k}$.
Clearly $\{z_{\Par,j}\}$ is a basis in $\mathrm{T}_{\sol_{\Par}}\Mf$.

We use the implicit function theorem for the map
$G:\Hone\times \setPar \mapsto \mathbb{R}^{2d+2}$, defined by
\begin{equation}
    G_j(\psi,\Par) := \rdot{\psi-\sol_\Par}{\J^{-1} z_{\Par,j}} \ \
      \forall j=1,...,2d+2.
\end{equation}
We verify that ($i$) $G\in \C{1}$, ($ii$) $G(\sol_{\Par_0},\Par_0) = 0$
for any $\Par_0 \in \setPar$ and ($iii$) $\left.\partial_\Par
G(\sol_{\Par_0},\Par)\right|_{\Par=\Par_0}$ is invertible.

$G$ is in $\C{1}$ in $\Par$, since so are $\sol_\Par$ and $z_{\Par,j}$, 
and $G$ is $\C{1}$ in $\psi$ since it is linear in $\psi$. 
So ($i$) follows.

($ii$) follows directly from the definition of $G$.

To prove ($iii$), we use \eqref{eq:dk} to compute, 
\begin{equation}
   \left. \partial_{k} G_j(\sol_{\Par_0},\Par)\right|_{\Par=\Par_0} =
   -\rdot{z_{\Par_0,k}}{\J^{-1} z_{\Par_0,j}} =
   -\rdot{z_{\freq_0,k}}{\J^{-1} z_{\freq_0,k}} ,
\end{equation}
where in the last equality we used that $\TB$ is symplectic (see
Remark~\ref{rem:uni} below).  
With the definition of $\iom$ in Section
\ref{sec:man} we find
\begin{equation}
   \left. \partial_\Par G(\sol_{\Par_0},\Par)\right|_{\Par=\Par_0} = 
        \left.-\iom\right|_{\{z_{\freq_{0},k}\}_{k=1}^{2d+2}}.
\end{equation}
Here we use the notation 
$\left.\iom\right|_{\{z_{\freq_{0},k}\}_{k=1}^{2d+2}}$
to denote the matrix of $\iom$ in 
the basis $\{z_{\freq_{0},k}\}_{k=1}^{2d+2}$. 
Thus $\left. \partial_\Par
G(\sol_{\Par_0},\Par)\right|_{\Par=\Par_0}$ is invertible for
all $\Par_{0}$ by Lemma~\ref{lem:Oinv}. 
This shows ($iii$).

With properties ($i$)--($iii$) 
the implicit function theorem implies that there exists a unique $\C{1}$  
map $\Par = \Par(\psi)$, satisfying $G(\psi,\Par(\psi))=0$ 
in a neighborhood, $\set{V}_{\Par_0}$, 
of $\sol_{\Par_0}$. 

Now take $\Par_0=\{0,0,0,\freq_0\}$, with $\freq_0\in I$. 
Denote $\Mm:=\{a,\vip,\gamma\}$ so that $\mathcal{S}_\Mm:=\TB$. 
Then for all $\Mt$ the map $\tilde{\Par}$,  
defined on $\mathcal{S}_{\Mt}\set{V}_{\Par_0}$, 
\begin{equation}
   \tilde{\Par}(\psi):= \Mt\circ \Par(\mathcal{S}_\Mm^{-1}\psi)
\end{equation}
where $\Mt\circ \{\Mm,\freq\}= \{\Mt\circ \Mm,\freq\}$
and $\Mt\circ\Mm$ is defined by $\mathcal{S}_{\Mt}\circ 
\mathcal{S}_\Mm=:\mathcal{S}_{\Mt\circ\Mm}$, solves equation
\eqref{eq:od3} for any $\psi\in \mathcal{S}_{\Mt}\set{V}_{\Par_0}$.
Since the neighborhood $ \mathcal{S}_{\Mt}\set{V}_{\Par_0}$
with $\Mt\in \mathbb{R}^d\times\mathbb{R}^d\times [0,2\pi)$
and $\freq_0\in I$ covers the neighborhood $U_\delta$ the statement of the
proposition follows. \qed
\end{proof}

So if we know that for a given initial condition \eqref{eq:sp1} has a
$\set{C}(\mathbb{R},\Hone(\mathbb{R}^d))\cap
\C{1}(\mathbb{R},\Sob{-1}(\mathbb{R}^{d}))$ solution $\psi=\psi(t)$
which, for times $0\leq t\leq T$, stays in the neighborhood
$U_\delta$, then by Proposition~\ref{prop:IFT} the solitary wave
solution parameters $\Par$ trace out a unique $\C{1}$ trajectory
$\Par(\psi(t))$. We make the choice $\epsE\ll \delta$.

\begin{remark}\label{rem:uni}
The operator $\TB$ is a symplectic, or canonical, operator; \ie it
leaves $\rdot{\cdot}{\J^{-1}\cdot}$, unchanged:
$\rdot{\TB w}{\J^{-1}\TB q}=\rdot{w}{\J^{-1}q}$. 
This follows from the fact that $[\J,\TB]=0$.
Since $z \in \set{T}_{\TB \solw}\Mf$ implies that
$\TB^{-1}z\in \set{T}_{\solw}\Mf$, we can write \eqref{eq:od3} in the
form
\begin{equation}
        \rdot{\TB^{-1}\psi-\solw}{\J^{-1} z} = 0 \; , \ 
         \forall z \in \set{T}_{\solw} \Mf \; .
         \label{eq:od4}
\end{equation}
where $\solw$ and $z$ depend on $\freq(t)$.
\end{remark}

\section{Equation of motion in the moving frame}
\label{sec:mf}

Given a solution $\psi$ to \eqref{eq:sp1}, we define the
parameterization $\{\Par,w\}$ for it by the equations
\begin{equation}\label{w}
\psi -\sol_{\Par} 
\bot \J^{-1}\set{T}_{\sol_{\Par}}\Mf\ \text{and}\ 
w:=\TB^{-1}(\psi-\sol_{\Par}).
\end{equation}
By Proposition~\ref{prop:IFT}, this parameterization is well defined
as long as $\psi \in U_{\delta}$.
In this section we find the equation for the parameters $\{\sigma,w\}$.
To this end we use the equation for the function $u$, defined by
\begin{equation}
   u := \TB^{-1}\psi.  \label{u}
\end{equation}
We introduce the anti-self-adjoint generators
\begin{equation}
   \lk_{j} = \partial_{x_j}, \ \  
   \lk_{d+j} = -\J x_j, \ \   \lk_{2d+1}=-\J,\ \
   \lk_{2d+2}=\partial_\freq,\ \ j=1,...,d 
   \label{eq:gen}
\end{equation}
and the corresponding coefficients
\begin{equation}
   \alpha_{j} = \dot{a}_j- \vip_j, \ \
   \alpha_{d+j} = -\frac{1}{2}\dot{\vip}_j - \partial_{x_j} V(a), \ \ 
   j=1,...,d,     \label{eq:mu}
\end{equation}
\begin{equation}
   \alpha_{2d+1}=\freq-\frac{1}{4}\vip^2+\frac{1}{2}\dot{a}\cdot \vip
        -V(a)-\dot{\gamma}, \ \
   \alpha_{2d+2}=-\dot{\freq}. \label{eq:nu}
\end{equation}
Let 
\begin{equation}
   \Lal = \sum_{j=1}^{2d + 2} \alpha_j \lk_j \ \ \mbox{and}\ \
   \Sal = \sum_{j=1}^{2d + 1} \alpha_j \lk_j.
\end{equation}
The main result in this section is
\begin{lemma}
   If $\psi$ satisfies \eqref{eq:sp1} then $u$ satisfies 
\begin{equation}
   \dot{u} = \J \big((-\Laplace +\freq)u - f(u) \big) + \Sal u +
   \J \VR u, \label{eq:ee104b}
\end{equation}
where 
\begin{equation}
   \VR:= V(x+a) - V(a) - \nabla V(a)\cdot x = 
        \Oh((\eps x)^2),  \label{eq:VR}
\end{equation}
\end{lemma}
\begin{proof}
For this proof, we will use complex notation, \ie $\J = \i^{-1}$,
and in particular we note that $[\J,\TB]=0$.  Let $\psi_a(x)=\psi(x+a)$,
and let $\phi=\frac{1}{2}\vip\cdot x + \gamma$.  With this notation
$u=\exp{-\i\phi}\psi_a$. Differentiating \eqref{u}, we find
\begin{equation}
   \i \dot{u} =  \exp{-\i \phi}(-\Laplace + \freq + V_a)\psi_a 
   - f(\psi_a) + 
   \i \exp{-\i \phi} \dot{a}\cdot \nabla \psi_a +
   \\
   (\frac{1}{2}\dot{\vip}\cdot x+ \dot{\gamma}-\freq)u \; ,
   \label{eq:sp7}
\end{equation}
where $V_a(x)=V(x+a)$ and $\exp{-\i\phi}f(\psi_a)=f(u)$, by 
Condition~\ref{con:g.sym}. We rewrite the
potential, using  \eqref{eq:VR}, in the form
\begin{equation}\label{eq:VVVV}
V(x+a)=V(a)+\nabla V(a)\cdot x+ \VR(x).
\end{equation}
Now use
\begin{equation}
   \exp{-\i \phi} \nabla \psi_a = \nabla(\exp{-\i \phi}\psi_a) +
     \i \exp{-\i \phi}\psi_a  \nabla{\phi}
\end{equation}
and
\begin{align}
   \exp{-\i \phi}\Laplace \psi_a &= \Laplace(\exp{-\i\phi}\psi_a) + 
    \i \nabla \phi \cdot \nabla (\exp{-\i \phi}\psi_a) + 
    \i \exp{-\i \phi}\nabla \phi \cdot \nabla \psi_a \\ 
        & = \Laplace(\exp{-\i\phi}\psi_a) + 
    2\i \nabla \phi \cdot \nabla(\exp{-\i\phi}\psi_a) 
    - |\nabla \phi|^2 \exp{-\i\phi}\psi_a \; ,
\end{align}
where $\phi = \frac{1}{2}\vip \cdot x + \gamma$, to conclude
\eqref{eq:ee104b}.  Note that the parameters and their time
derivatives, as well as $V(a)$ and $\nabla V(a)$ are collected into
$\Ass$.\qed
\end{proof}

We now re-parameterize the non-linear Schr\"odinger equation into
separate equations for $\Par$ and $w$. 
Recalling that $u=\sol+w$, $\partial_t \sol = \dot{\freq}\partial_\freq \sol$,
$\Ew'(\sol)=0$, and $\Ew'(\sol+w) = \LL w + \NII{w}$, 
where $\LL=\Ew''(\sol)$,
we can rewrite \eqref{eq:ee104b} as
\begin{equation}\label{eq:mellan}
\dot{\freq} \partial_\freq \sol +  \dot{w} = \J\LL w + \J \NII{w} + 
\Sal (\sol+w) +\J \VR (\sol+w)
\end{equation}
We collect the linear terms acting on $w$ as $\Le w$, where
\begin{equation}\label{eq:Le}
\Le := \LL + \VR + \J^{-1}\Sal,
\end{equation}
and the remaining terms into a source term
\begin{equation}\label{eq:f}
q(\Par) := \Lal \sol + \J\VR\sol,
\end{equation}
to obtain
\begin{equation}
\dot{w} = \J \Le  w+ \NII{w} + q(\Par).
\end{equation}

The equations for the parameters are obtained using the 
skew orthogonality condition and \eqref{eq:mellan}.
Let $z\in \TM$.
Upon recalling that $\dotp{\J z}{\J \LL w}=0$, and so
\begin{equation}
0=\partial_t \dotp{\J z}{w}= 
\dotp{\J z}{\dot{w}} + 
\dot{\freq}\dotp{\J \partial_\freq z}{w},
\end{equation}
we find
\begin{equation}\nonumber
   \dot{\freq}\dotp{\J z}{\partial_\freq \sol} - \Ass \cdot \dotp{\J
     z}{\lss \sol} = \dot{\freq}\dotp{\J \partial_\freq z}{w}+
   \dotp{z}{\NII{w}}+\Ass \cdot \dotp{\J z}{\lss w} + \dotp{z}{\VR
     (\sol+w)}.
\end{equation}
Recall that $\partial_\freq \sol = \lk_{2d+2}\sol$, so the left-hand
side is $\sum_j \dotp{\J z}{\lk_j\sol}\alpha_j=\sum_j \dotp{\J
z}{z_j}\alpha_j$, where we used $\lk_j\sol=z_j$. Now let $z$ be
one of the basis vectors, $z=z_k$. Then the inner product on the
left-hand side coincides with the definition of
$(\iom)_{kj}$. Furthermore, using $\adjoint{\lss}=-\lss$, and
$[\lss,\J]=0$, we combine $\Ass\cdot\dotp{\J z}{\lss w}$ and and
$\dot{\freq}\dotp{\J\partial_\freq z}{w}$, into $\alpha \cdot \dotp{\lk z}{\J
w}$. The result is
\begin{equation}\label{eq:inter}
\sum_{j=1}^{2d+2}(\Omega^{-1})_{kj}\alpha_j = \dotp{z_k}{\NII{w}+\VR (w+\sol)}
+ \alpha\cdot \dotp{\lk z_k}{\J w} .
\end{equation} 
Replacing $\alpha$ by the explicit expression, 
\eqref{eq:inter} reads, for $k=2d+1$ and $k=1,..,d$,
\begin{align}
\dot{\freq}&=(\mm'(\freq))^{-1} \left(
\dotp{\sol}{\J \NII{w}+\J\VR w} - \alpha\cdot\dotp{\lk
\sol}{w}\right), \\ 
\frac{1}{2}\dot{v}_k & = -\partial_{x_k}V(a) + (\mm(\freq))^{-1}\left(
\dotp{\partial_k\sol}{\NII{w}+\VR w}-
\alpha\cdot\dotp{\lk \partial_k \sol}{\J w} \right. \nonumber \\
& \quad \left. +\dotp{\partial_k \sol}{\VR \sol} \right),
\end{align}
where $\mm(\freq):=2^{-1}\nrm{\sol}^{2}$, and we have 
used $\dotp{\J \sol}{\VR\sol}=0$.
For $k=d+1,...,2d$, $k=2d+2$, we use the expressions for $\dot{\freq}$ 
and $\dot{v}_k$ obtained above to find 
\begin{align}
\dot{a}_k&=\vip_k+ (\mm(\freq))^{-1}\left[\dotp{x_k \sol}{\J \NII{w}+\J \VR
w}+\alpha\cdot\dotp{\lk x_k \sol}{w}
\right] ,
\\ 
\dot{\gamma}
& = \freq-\frac{1}{4}\vip^2+\frac{1}{2}\dot{a}\cdot \vip- V(a)- 
(\mm'(\freq))^{-1}\left[\dotp{\partial_\freq \sol}{\NII{w}+ \VR w}  
\right. \nonumber \\ & \left. \quad 
- \alpha\cdot\dotp{\lk \partial_\freq \sol}{\J w}  
 + \dotp{\partial_\freq \sol}{\VR \sol}
\right], \nonumber
\end{align}
where we used $\dotp{x_k \sol}{\J \VR \sol}=0$, and  
$\sol(x)=\sol(|x|)$ so that $\dotp{x\sol}{\sol}=0$.
Furthermore, observe that
all terms containing $w$ and $\VR$ are of higher order. 
We abbreviate this as follows 
\begin{equation}\label{eq:end}
\dot{\Par} = X(\Par) - \delta X(\Par, w). 
\end{equation}
For the estimates used later, we note that $\dot{\Par_j}-X_j(\Par)=\alpha_j$.
We formalize the above calculation in the following Proposition:
\begin{proposition}\label{prop:aw}
(1) The parameters $\Par$ 
and the fluctuation $w$ (defined in \eqref{w}) satisfy the
equations
\begin{equation}
   \dot{\Par}= X(\Par)-\delta X(\Par,w), \label{eq:Aj} 
\end{equation}
and
\begin{equation}
    \dot{w} = \J \Le w 
          + \J \NII{w}+ q(\Par) \label{eq:dw}.
\end{equation}
Here $\Le$, $q(\Par)$ and $\NII{w}$ are 
given by \eqref{eq:Le}, \eqref{eq:f} and 
\eqref{eq:N} respectively, 
and (with $\om$ defined in Lemma~\ref{lem:Oinv})
\begin{equation}\label{eq:dX}
   \delta X_j(\Par ,w)=\sum_{k=1}^{2d+2}(\om)_{jk}
   [\rdot{z_k}{ \NII{w}+\VR (w+\sol)}+\alpha\cdot \dotp{\lk z_k}{\J w }],
\end{equation}
$\forall j=1,...,2d+2$, with $\{z_k\}_{k=1}^{2d+2}:=\{\tr, \ga,\gu, 
\sa\}$, and $\VR$ given by \eqref{eq:VR}.

(2) The vector field $\delta X$ satisfies the 
following estimate for $\nrmHo{w}\leq 1$:
\begin{equation}
    \delta X = \Oh(\anrm{\alpha}\nrm{w} + \eps^2 + 
        \nrmHo{w}^2) \; , \label{eq:OhX}
\end{equation}
where $\anrm{\alpha} := \max_{j=1,...,2d+2} |\alpha_j|$.
\end{proposition}

\begin{proof}
(1) is the result of the calculation done in
\eqref{eq:mellan}--\eqref{eq:end}. In particular \eqref{eq:dX} follows from
\eqref{eq:inter}.

(2) Estimate \eqref{eq:OhX} follows directly from \eqref{eq:dX}, 
together with the facts  
that 
$\nrm{\VR z_k}=\Oh(\eps^2)$ and $\nrmD{\NII{w}} \leq c\nrmHo{w}^2$ for 
$\nrmHo{w}\leq 1$ (see \eqref{eq:N}).\qed
\end{proof}

The goal is to show that $\sup_{t\in (0,T)}|\delta X|=\Oh(\eps^2+\epsE^2)$ and 
$\sup_{t\in(0,T)} \nrmHo{w}=\Oh(\eps+\epsE)$, 
for some 
$T=\Oh(1/(\eps+\epsE^{2}))$.

\section{Approximate Conservation of a Lyapunov functional}
\label{sec:Lf}

In this section, we show that the Lyapunov functional
$\Ew(u)-\Ew(\sol_{\freq})$, is approximately conserved. Recall $\sol=\solw$ is
the solitary wave profile (see Section~\ref{sec:intro})
and $u$ is the solution
$\psi$ of \eref{eq:sp1}, transformed to the moving frame: 
$u:=\TB^{-1}\psi$
($\TB$ is defined in Section~\ref{sec:man}).
We use the skew-orthogonal
decomposition of $u$ (see Section~\ref{sec:on}):
$u:=\TB^{-1}\psi = \sol + w$, with 
\begin{equation}\label{eq:skew}
  \stfrm_\sol(w,z)=0\ \ \text{for all}\ z\in \TM,  
\end{equation}
provided $\psi\in U_\delta$. The main result of this section
is the following
\begin{proposition}\label{prop:EE}
Let $\psi\in U_\delta$ 
solve \eref{eq:sp1} and let $u$, $w$ and $\sol$ be defined as
above. Then
\begin{equation}\label{eq:EE}
\partial_t (\Ew(u)-\Ew(\solw)) = \Oh(|\alpha|\nrmHo{w}^2 + 
\eps^2 \nrmHo{w}+\eps \nrmHo{w}^2).
\end{equation}
\end{proposition}
To prove this proposition we use the following
\begin{lemma}\label{lem:tEw} Let $u$ be defined as above. Then 
\begin{equation}
\partial_t \Ew(u) = \frac{1}{2}\dot{\freq}\nrm{u}^2
 -\dotp{(\frac{1}{2}\dot{\vip}+\nabla V_a)\i u}{\nabla u}.
\end{equation}
\end{lemma}
\begin{proof}[of Proposition~\ref{prop:EE}]
We first recall that $\sol$ is a critical point to $\Ew(u)$, thus
\begin{equation}
\partial_t \Ew(\sol) = \frac{1}{2}\dot{\freq}\nrm{\sol}^2.
\end{equation}
Using Lemma~\ref{lem:tEw}, we find
\begin{equation}\label{eq:estEE}
\partial_t (\Ew(u)-\Ew(\sol)) = A - B.
\end{equation}
where $A:=2^{-1}\dot{\freq}(\nrm{u}^2-\nrm{\sol}^2)$ and 
$B:=\dotp{(2^{-1}\dot{\vip}+\nabla V_a)\i u}{\nabla u}$.
First we use the decomposition $u=\sol+w$,  
the condition   
$0=\dotp{\i \gu}{w}=\dotp{\sol}{w}$ (from \eqref{eq:skew}),
and the estimate
$|\dot{\freq}|=|\alpha_{2d+2}| \leq \anrm{\alpha}$ to obtain
\begin{equation}\label{eq:A}
A=\frac{1}{2}\dot{\freq}\nrm{w}^2 = \Oh(|\alpha|\nrm{w}^2).
\end{equation}

For the term $B$, recall that $0=\dotp{\i \tr}{w}=\dotp{\i \nabla\sol}{w}$,
and furthermore that $\dotp{\i q\sol}{\nabla \sol}=0$ 
for any real-valued function $q\in \Lp{\infty}$. Then
\begin{equation}
B = \dotp{(\frac{1}{2}\dot{\vip}+\nabla V_a)\i w}{\nabla w} + 
\dotp{(\nabla V_a)\i \sol}{\nabla w}+\dotp{(\nabla V_a)\i w}{\nabla \sol}.
\end{equation}
Now, we use that $\nabla V(a)\cdot \dotp{\i w}{\nabla \sol}=0=\nabla V(a)
\cdot\dotp{\i \sol}{\nabla w}$, to obtain 
\begin{multline}
B = (\frac{1}{2}\dot{\vip}+\nabla V(a))\cdot\dotp{\i
w}{\nabla w} + \dotp{(\nabla V_a - \nabla V(a))\i
w}{\nabla w}+\\ \dotp{(\nabla V_a-\nabla V(a))\i \sol}{\nabla w}+\dotp{(\nabla
V_a-\nabla V(a))\i w}{\nabla \sol}.
\end{multline}
Recall that $-\alpha_{d+j}:=\frac{1}{2}\dot{\vip}_j+\partial_j V(a)$ (see
\eqref{eq:nu}). Since $|\alpha_j|\leq\anrm{\alpha}$, the first term on
the right hand side is $\Oh(\anrm{\alpha}\nrmHo{w}^2)$. Since $\nabla
V_a = \Oh(\eps)$ the second term on the right-hand side is
$\Oh(\eps\nrmHo{w}^2)$. Finally due to $\nabla V_a-\nabla
V(a)=\Oh(\eps^2|x|)$ and $|x|\sol, |x|\nabla \sol\in \Ltwo$, 
(Condition~\ref{con:g.Omega}) the third
and fourth terms are $\Oh(\eps^2\nrmHo{w})$. Collecting these estimates
we arrive at
\begin{equation}\label{eq:B}
B = \Oh(\anrm{\alpha}\nrmHo{w}^2 + \eps^2 \nrmHo{w}+\eps \nrmHo{w}^2).
\end{equation}
Relations \eqref{eq:estEE}, \eqref{eq:A} and \eqref{eq:B} imply \eqref{eq:EE}.
\qed
\end{proof}
It remains to prove Lemma~\ref{lem:tEw}. To this end we note the following
\begin{lemma}\label{lem:REL} Let $\psi$ be a solution to \eref{eq:sp1}. 
Then
\begin{equation}\label{eq:btr}
\partial_t \frac{1}{2}\int V |\psi|^2 = \dotp{(\nabla V)\i \psi}{\nabla\psi}.
\end{equation}
\end{lemma}
\begin{proof}
\eqref{eq:btr} is obtained by integrating the relation
\begin{equation}
\partial_t (V|\psi|^2) =\i \nabla \cdot (V\bar\psi \nabla \psi -V \psi
\nabla \bar\psi) - \i (\nabla V)\cdot(\bar\psi\nabla\psi-\psi\nabla\bar\psi),
\end{equation}
which follows from the nonlinear Schr\"odinger equation \eqref{eq:sp1}.
\qed
\end{proof}

\begin{proof}[of Lemma~\ref{lem:tEw}]
Note that the identity 
\begin{equation}\label{eq:eid}
\Ham_V(\TB^{-1}\psi) + 
\frac{1}{2}\freq \nrm{\TB^{-1}\psi}^2 - \frac{1}{2}\int V
|\TB^{-1}\psi|^2 = \Ew(u).
\end{equation}
holds for all $u=\TB^{-1}\psi$.
We observe the following relations
\begin{equation}
\nrm{\TB^{-1}\psi}^2 = \nrm{\psi}^2, \ \ \int V
|\TB^{-1}\psi|^2 = \int V_{-a} |\psi|^2
\end{equation}
and
\begin{equation}
2\Ham_V(\TB^{-1}\psi) = 2\Ham_V(\psi) + \frac{1}{4}\vip^2 \nrm{\psi}^2 -
\vip \cdot \dotp{\i \psi}{\nabla \psi} + \int (V_{-a} -V)|\psi|^2 
\end{equation}
Inserting the above relations into \eqref{eq:eid} we find
\begin{equation}
2\Ham_V(\psi)+ (\frac{1}{4}\vip^2 + \freq) \nrm{\psi}^2-
\vip \cdot \dotp{\i
  \psi}{\nabla \psi} - \int V |\psi|^2 = 2\Ew(u).
\end{equation}

We now take the time derivative of the above relation. Using 
the fact that
$\Ham_V(\psi)$ and $\nrm{\psi}^2$ are conserved quantities, and using 
Lemma~\ref{lem:REL} and Ehrenfest's theorem \eqref{eq:Ehrenfest} 
we find
\begin{equation}
(\frac{1}{2}\dot{\vip}\cdot \vip+\dot{\freq})\nrm{\psi}^2 -
\dot{\vip}\cdot\dotp{\i\psi}{\nabla \psi} 
+\vip\cdot\dotp{\nabla V\psi}{\psi} -
2\dotp{(\nabla V)\i \psi}{\nabla \psi}=2\partial_t \Ew(u).
\end{equation}
Collecting terms of the form $(\frac{1}{2}\dot{\vip}+\nabla V)$ 
and replacing $\psi$ with  $\TB u$ gives 
\begin{equation}
\dot{\freq}\nrm{u}^2+
\dotp{\i (\dot{\vip}+2\nabla V_a)\cdot \nabla u}{u} 
=2\partial_t \Ew(u).
\end{equation}
By observing that $\dotp{q\i\nabla u}{u}=-\dotp{q\i u}{\nabla
u}$ for real $q\in\Lp{\infty}$, we arrive at the
result of the lemma.
\qed
\end{proof}

\section{Lower bound on the Lyapunov functional}
\label{sec:lL}

\begin{proposition}\label{prop:Ebbl}
Let $\sol$ and $w$ be as in Proposition~\ref{prop:EE}.
Then there exist constants $\rho>0$ and $c>0$ independent of 
$\eps$ and $\epsE$ such that for $\nrmHo{w}\leq 1$,
\begin{equation}\label{eq:Ebbl}
   |\Ew(\sol+w) - \Ew(\sol)| \geq  \frac{\rho}{2}\nrmHo{w}^2 - c
\nrmHo{w}^3 \; .
\end{equation}
\end{proposition}
\begin{proof}
We expand $\Ew(u)$ around $\sol$. Using
that $\sol$ is a critical point to $\Ew$ ($\Ew'(\sol)=0$), we write
\begin{equation}\label{eq:exp}
   \Ew(\sol+w) - \Ew(\sol) = \frac{1}{2} \rdot{w}{\LL w}
   + \RIII{w} \; ,
\end{equation}
where, recall, $\LL:= \Ew''(\sol)$ and where 
$\RIII{w}$ is defined in \eqref{eq:RIII}. From 
condition~\ref{con:g.ham} and \eqref{R} we have for 
$\nrmHo{w}\leq 1$
\begin{equation}
    |\RIII{w}|\leq c \nrmHo{w}^3 \; .
    \label{eq:R}
\end{equation}

Let $\set{Y}_\sol:=\{w\in\Hone(\mathbb{R}^d): 
\stfrm_\eta(w,z)=0, \forall z\in \TM, \nrmHo{w}=1\}$. 
It is shown in Appendix \ref{app:BBL} (see \cite{Weinstein1986}) 
that 
\begin{equation}\label{eq:rho-eq}
\rho:=\inf_{w\in \set{Y}_\sol}\dotp{w}{\LL w}>0.
\end{equation}
Hence, for $w$ that satisfy \eqref{eq:skew} we have the following
coercivity estimate
\begin{equation}\label{eq:rhoX}
\dotp{w}{\LL w} \geq \rho \nrmHo{w}^2.
\end{equation}
Using this
estimate and the bound \eqref{eq:R} on $R_{\sol}$ in \eqref{eq:exp} we
arrive at \eqref{eq:Ebbl}.  \qed
\end{proof}
\begin{remark}
Since
$\sigma_{\mathrm{ess}}(\LL)=[\freq,\infty)$, we have that $\rho\leq
\freq$. We expect that for a wide class of nonlinearities 
$\rho\geq c\freq$ for some constant $c>0$.
\end{remark}

\section{Upper bound on $\nrmHo{w}$ and proof of the main result}
\label{sec:MT}

In this section we prove the main result of the paper, by providing
an upper bound on $\nrmHo{w}$.  To achieve this, we use
both the approximate conservation of $\Ew$ and the lower bound on the
Lyapunov functional.  For vector functions $s\mapsto w(s)\in \Hone$,
and $s\mapsto \alpha(s)\in \mathbb{R}^{2d+2}$ we introduce the norms
$\WW := \sup_{s\in [0,t]} \nrmHo{w(s)}$ and $\Anrm{\alpha} :=
\sup_{s\leq t}\anrm{\alpha(s)}$.  
We state the main result of this section

\begin{proposition}\label{prop:wup}
Assume $\eps$, $\epsE$ are sufficiently small. 
There are constants $c,c'<\infty$, independent of $\eps$ and $\epsE$ 
such that for $t\leq c(\eps+ \epsE^2)^{-1}$,
\begin{align}
   \nrmHo{w}&\leq c'(\eps + \epsE), 
   \label{eq:wupA}\\ 
   \Anrm{\alpha} &\leq c'(\eps^2 + \epsE^2), \label{eq:wupB}
\end{align}
where $w=\TB^{-1}\psi-\solw$ and $\alpha_j=\dot\Par_j-X_j(\sigma)$ 
(see \eqref{eq:Aj}).
\end{proposition}
Denote 
\begin{equation}\label{eq:tri}
   \triangle \En := \En_{\freq_0}(u_0)-\En_{\freq_0}(\sol_{\freq_0})
   \; ,
\end{equation}
where $u_0:=\left.\TB^{-1}\psi\right|_{t=0}$.
We begin with two simple auxiliary lemmas.
\begin{lemma}\label{lem:dE} There exists a constant $c>0$ independent 
of $\eps$ and $\epsE$ such that 
$|\triangle \En|$ satisfies the inequality
\begin{equation}\label{est:DE}
|\triangle \En|\leq c \epsE^2.
\end{equation}
\end{lemma}
\begin{proof}
To estimate $\triangle \En$, we use the fact that 
$\DE_{\freq_0}(\sol_{\freq_0}) = 0$ to obtain
\begin{align}
     2\triangle \En &= 2(\En_{\freq_0}(\sol_{\freq_0} + w_0) 
        - \En_{\freq_0}(\sol_{\freq_0})
        - \rdot{\DE_{\freq_0}(\sol_{\freq_0})}{w_0}) \nonumber \\& = 
        \nrm{\nabla w_0}^2  + \freq_{0}\nrm{w_0}^2 -\RII{w_{0}}. 
     \label{eq:pmt1}
\end{align}
Since $\nrmHo{w_0}\leq 1$, the estimate  
\eqref{R} gives
\begin{equation}
    \nrm{\RII{w_0}} \leq c \nrmHo{w_0}^{2} \; .
\end{equation}
This together with the identity \eqref{eq:pmt1} and the fact that 
$\nrmHo{w_{0}}\leq \epsE<1$ gives the estimate 
\eqref{est:DE}.\qed
\end{proof}
\begin{lemma}\label{lem:wb1}
    Let $\rho>0$ be the coercivity constant given in \eqref{eq:rho-eq}. 
There exists a constant $c$ independent of 
    $\eps$ and $\epsE$, such that for $\nrmHo{w}\leq 1$
\begin{equation}\label{eq:L7}
    \rho \WW^{2} \leq c\epsE^{2} + 
    ct(\eps^{2}\WW+(\eps+\Anrm{\alpha})\WW^{2}) + c\WW^{3}
\end{equation}
\end{lemma}
\begin{proof}
Using the approximate 
conservation of the Lyapunov functional (the time integral of 
\eqref{eq:EE}),
and  pulling $\sup$ of norms 
out of the time integral we obtain
\begin{equation}
      |\Ew(\solw+w) - \Ew(\solw)| \leq |\triangle \En|+
 c t (\eps^2 \WW + (\eps+\Anrm{\alpha}) \WW^2)
      \label{eq:Ettt}
\end{equation}
where $|\triangle \En|$ is defined in \eqref{eq:tri}.
Now we substitute
\eqref{eq:Ettt} into the lower bound \eqref{eq:Ebbl}, use 
the initial condition estimate \eqref{est:DE} and take the 
$\sup_{s\in[0,t]}$ of the resulting expression to obtain
\eqref{eq:L7}.\qed
\end{proof}
\begin{proof}[of Proposition~\ref{prop:wup}]
Using the triangle inequality we derive from \eqref{eq:L7} that the 
function $X(t):=\WW$ satisfies the equation
\begin{equation}\label{e}
    0\leq \rho \eps^{2}+ c \epsE^{2} - \rho X^{2}+ c X^{3}
\end{equation}
for times $c t (\eps + |\alpha|)\leq \rho/4$, provided $X\leq 1$. 
The graph of the right-hand side of \eqref{e} is shown in Figure 
\ref{fig:tt}.
\psfrag{X}{$X$}
\psfrag{p}{$p$}
\psfrag{q}{$q$}
\psfrag{Q}{$q_0$}
\psfrag{1}{$1$}
\begin{figure}
   \centering
\includegraphics{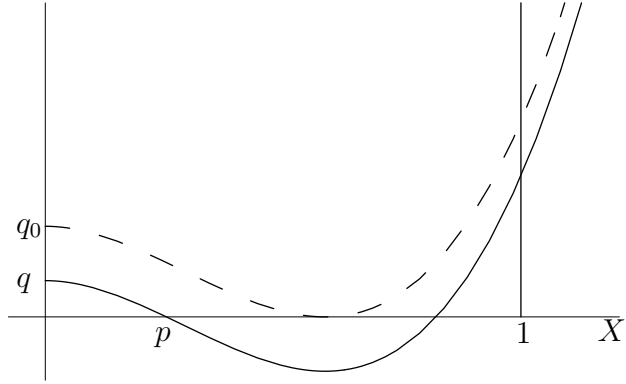}
   \caption{\small Schematic graph of the left-hand side of \eqref{e}.}
        \label{fig:tt} 
\end{figure}
Thus for $\rho \eps^{2}+c \epsE^{2}<q_0$, 
see Figure~\ref{fig:tt}, $X=\WW\leq p$ if
$\left. X\right|_{t=0}\leq p$, where $p$ is the smallest positive zero
of the left-hand side of \eqref{e} (see Figure~\ref{fig:tt}), provided 
$p\leq 1$. For
$\epsE+\eps$ sufficiently small,
$c'(\eps+\rho^{-1/2}\epsE)\leq p \leq c(\eps+\rho^{-1/2}\epsE)$ 
and therefore
\begin{equation}
   \nrmHo{w}\leq \WW \leq c(\eps + \frac{1}{\sqrt{\rho}}\epsE) \label{eq:WWb}
\end{equation} 
provided $\nrmHo{w_0}\leq 1$. 
Substituting \eqref{eq:WWb} into \eqref{eq:OhX} we obtain from \eqref{eq:Aj}
and the notation $\alpha:=\dot\Par-X(\Par)$ that 
\begin{equation}\label{eq:AAE}
    \Anrm{\alpha}\leq c(\eps^{2}+\rho^{-1}\epsE^2),
\end{equation}
for times $t\leq c (\eps + \rho^{-1}\epsE^2)^{-1}$. This completes the proof.\qed
\end{proof}

\appendix
\fixNumberingInAppendix

\section{Ehrenfest's theorem}\label{app:Ehrenfest}

In this appendix we prove Ehrenfest's theorem, 
\eref{eq:Ehrenfest}. Denote $\Ip{\psi}(t):=\dotp{\psi}{-\i 
\partial_{x_{j}}\psi}$. Let $\psi$ solve eq.~\eqref{eq:sp1}
and assume that 
$\psi\in\set{C}(\mathbb{R},\Htwo)\cap\C{1}(\mathbb{R},\Ltwo)$. 
Then $\Ip{\psi}(t)$ is in $\C{1}$,
and after a straightforward calculation using \eref{eq:sp1}, we find
\begin{equation}
\dIp{\psi} = \int - |\psi|^2\partial_{x_j} V\diff^d x =:L_\psi(t).
\end{equation}
By the fundamental theorem of calculus, we obtain
\begin{equation}
\Ip{\psi}(t) = \Ip{\psi}(t_0) + \int_{t_0}^t L_\psi(s)\diff s.
\end{equation}

Now take an initial condition $\psi_{0}\in\Hone$. 
Pick 
$\psi_{0,n}\in\Htwo$ such that $\psi_{0,n}\rightarrow \psi_{0}$ in $\Hone$. 
Then by Theorems 5.2 and 4.2
of \cite{Cazenave1996}, the solutions $\psi_{n}$
corresponding to the initial conditions $\psi_{0,n}$ satisfy $\psi_{n}\in 
\set{C}(\mathbb{R},\Htwo)\cap \C{1}(\mathbb{R},\Ltwo)$ and 
$\psi_{n}\rightarrow\psi$ in $\set{C}(\mathbb{R},\Hone)$. 
Thus since
\begin{equation}
I_{\psi_n}(t)=I_{\psi_n}(t_0) + \int_{t_0}^{t} L_{\psi_n}\diff s,
\end{equation}
we have
\begin{equation}\label{eq:IIL}
I_{\psi}(t)=I_{\psi}(t_0) + \int_{t_0}^{t} L_{\psi}\diff s.
\end{equation}
We furthermore observe that both $\Ip{\psi}(t)$ and $L_{\psi}(t)$ are 
continuous in $t$ for 
$\psi\in\set{C}(\mathbb{R},\Hone)\cap\C{1}(\mathbb{R},\Sob{-1})$. 
Hence \eqref{eq:IIL} implies that 
$\Ip{\psi}(t)$ is $\C{1}$
and satisfies $\dIp{\psi}(t)=L_{\psi}(t)$.
This proves \eqref{eq:Ehrenfest}.\qed

\section{Minimization under 
constraint and spectrum of Hessian}
\label{app:spec}

In this appendix we show that the operator $\LL:=\Ew''(\sol)$ has
exactly one negative eigenvalue. 
The argument below is probably well-known,
but we did not find it in the literature.
Let $\mathrm{X}$ be a Banach space and $K\in \C{3}(\mathrm{X},
\mathbb{R})$ be a given 
functional. Define the set
\begin{equation}\label{eq:C1}
        \Man = \{u \in \mathrm{X}: K(u)=0\} \; .
\end{equation}
We have the following
\begin{proposition}\label{prop:neg}
Let $\En$ be a $\C{2}$ functional on $\mathrm{X}$.  Assume there is a
Hilbert space, $\set{H}$, such that $\set{H}\supset \set{X}$, densely,
and that the Hessian quadratic form $\Hess\En(u)(\alpha,\beta)$,
$\alpha, \beta\in \set{X}$, defines a self-adjoint operator $\En''(u)$
on $\set{H}$ such that
$\dotp{\alpha}{\En''(u)\beta}=\Hess\En(u)(\alpha,\beta)$, $\forall
\alpha \in \set{X}$, $\beta\in \set{D}(\En''(u))\subset \set{X}$.  Let
$\sol$ be a minimizer of $\En$ on the set $\Man\subset{\mathrm{X}}$
defined in \eqref{eq:C1}.  Assume $K'(\sol)\neq 0$.  Then the Hessian
operator $\En''(\sol)$ has at most one negative eigenvalue.
\end{proposition}
\begin{proof} Let $\sol$ be a minimizer of $\En$ on $\Man$. 
Then $\sol$ satisfies
\begin{equation}
        \En'(\sol)=0 \ \text{and} \ \En''(\sol)\geq 0 \; . 
        \label{eq:crt}
\end{equation}
Here $\En'(\sol): \TMan{\sol}\mapsto 
\mathbb{R}$ and $\En''(\sol): \TMan{\sol}\mapsto (\TMan{\sol})^*$, where
\begin{equation}\label{eq:C3}
        \TMan{\sol} = \{\left. \partial_{s} 
        \sol_{s}\right|_{s=0}:\sol_{s}\in \C{1}([0,\epsilon],\Man),\ \sol_{s=0}=\sol\} \; .
\end{equation}

We claim that 
$\TMan{\sol}$ can be written as
\begin{equation}
        \TMan{\sol} = \{\xi\in \set{X} : \dotp{K'(\sol)}{\xi}=0\} =:
        K'(\sol)^{\bot} \; .
\end{equation}
Indeed, if $\xi\in\TMan{\sol}$ then there exists $\sol_s$ as in \eqref{eq:C3}
with $\xi=\left.\partial_s\right|_{s=0}\sol_s$ and therefore
\begin{equation}
        0 = \left. \partial_{s} K(\sol_{s}) \right|_{s=0} = \dotp{K'(\sol)}{\xi} 
        \; .
\end{equation}
On the other hand if $\xi\in K'(\sol)^\bot$ then we can find $\sol_s$
such that $\sol_{s=0}=\sol$, $\left.\partial_s\right|_{s=0}\sol_s =\xi$ and 
$K(\sol_s)=0$ by solving the equation $f(a,s)=0$ where
\begin{equation}
f(a,s):=\frac{1}{s^2}K(\sol+s\xi+s^2 a K'(\sol))
\end{equation}
for $a$ and setting $\sol_s=\sol+s\xi+s^2 a K'(\sol)$. 
The latter equation 
has a unique solution for $s$ sufficiently small since $f(b,0)=0$, where
$b=-\dotp{\xi}{K''(\sol)\xi}/\nrm{K'(\sol)}^2$ and 
$\partial_a f(b,0)=\nrm{K'(\sol)}^2$, and therefore the implicit function theorem is applicable.

Now, the second equation in 
\eqref{eq:crt} can be rewritten as
\begin{equation}
        \inf_{\xi \in K'(\sol)^{\bot}} 
        \dotp{\xi}{\En''(\sol)\xi} \geq 0.
\end{equation}
According to the max-min principle, the number of non-positive eigenvalues of 
$\En''(\sol)$ is less or equal to the co-dimension of 
$K'(\sol)^{\bot}$, which is 1.\qed
\end{proof}

In our case $\set{X}=\Hone(\mathbb{R})$, $\set{H}=\Ltwo(\mathbb{R}^d)$, 
$\En:=\Ew$, where 
\begin{equation}
        \Ew(u) 
        = \frac{1}{2} \int |\nabla u|^2 + \freq |u|^2 \diff^d x - 
        F(\psi) \; ,
\end{equation}
and $\sol$ is a minimizer to $\Ew$ with the constraint
\begin{equation}
        K(u) := \frac{1}{2} \int |u|^2 \diff^d x - m =0\; .
\end{equation}
This implies that $\LL:=\Ew''(\sol)$ has at most one negative
eigenvalue.

\begin{proposition}
Under condition~\ref{con:g.stab} $\LL$ has exactly one negative eigenvalue.
\end{proposition}
\begin{proof}
By Proposition~\ref{prop:neg} has at most one negative eigenvalue. One the 
other hand, since $\LL \partial_\freq \sol=-\sol$, we have 
\begin{equation}
   \rdot{\partial_\freq \sol}{\LL \partial_\freq \sol} = 
        - \frac{1}{2}\partial_\freq \int \sol^2\diff^d x <0,
\end{equation}
by Condition~\ref{con:g.stab}. Therefore by a variational principle
$\LL$ has at least one negative eigenvalue. Thus $\LL$ has exactly one
negative eigenvalue.\qed
\end{proof}

\section{The null space of $\LL$}
\label{app:nullL}

In this appendix we discuss condition~\ref{con:g.spec} (see
\eqref{eq:NullL}). We represent
complex functions $u(x)=u_1(x)+\i u_2(x)$ as real vectors
$(u_1(x),u_2(x))$.
In this representation the operator
$\LL$ takes the form
\begin{equation}
   \LL = \begin{pmatrix} L_1 & 0 \\ 0 & L_2 \end{pmatrix} \label{eq:Lm}\; ,
\end{equation}
(on $\Ltwo(\mathbb{R}^d,\mathbb{R})\oplus\Ltwo(\mathbb{R}^d,\mathbb{R})$),
where 
\begin{equation}
   L_1 = -\Laplace + \freq - f^{(1)}(\sol) \; , 
\end{equation}
and 
\begin{equation}
   L_2 = - \Laplace + \freq - f^{(2)}(\sol),
\end{equation}
with $f^{(1)}(\sol):=\partial_{\RE \psi}(\RE f)(\sol)$ and 
$f^{(2)}(\sol):=\partial_{\IM \psi} (\IM f)(\sol)$. The diagonal form 
of $\LL$ follows from the diagonal form of $f'(\sol)$: 
\begin{equation}
    f'(\sol)=\diag(f^{(1)}(\sol),f^{(2)}(\sol)).
\end{equation}
The latter follows from the relation $f(\CC\psi)=\CC f(\psi)$, where
$\CC$ is complex conjugation, and the fact that $\sol$ is real. 
This relation, in turn, follows from 
$F(\CC\psi)=F(\psi)$
(see Condition~\ref{con:g.sym}).

The matrix operator \eqref{eq:Lm} is then extended to 
$\Ltwo(\mathbb{R}^d,\mathbb{C})\oplus\Ltwo(\mathbb{R}^d,\mathbb{C})$.
The operators $L_1$ and $L_2$ are self-adjoint, with 
essential spectra given by 
\begin{equation}
   \sigma_{\mathrm{ess}}(L_j)=[\freq,\infty), \ \ j=1,2.
\end{equation}

Relation \eqref{eq:nulls} implies that 
\begin{equation}
   \partial_{x_j}\sol \in \Null{L_1}, \ \forall j=1,...,d,
\end{equation}
and 
\begin{equation}
   \sol\in \Null{L_2}.
\end{equation}

From now on, we assume that $f$ is a local nonlinearity. The 
fact that
$\sol>0$, implies by a Perron-Frobenius argument
(see \cite{Reed+Simon+4}) that 
\begin{equation}
   L_2\geq 0\ \text{and}\ \Null{L_2}= \mathbb{C}\sol. \label{eq:nL2}
\end{equation}

Thus it remains to analyze the operator $L_1$. To begin with, we
observe that since $\LL$ has exactly one negative
eigenvalue, relation \eqref{eq:nL2} implies
that $L_1$ has exactly one negative eigenvalue (the
same as $\LL$).

First we consider the case $d=1$. Then the zero mode, $\sol'$, of $L_1$
has exactly one zero (at $x=0)$ and consequently (by
Sturm-Liouville theory) $L_1$ has exactly one negative eigenvalue,
as we concluded above from general considerations. (Remember that 
the lowest eigenvalue in our case is simple, and that the 
nonlinearity is local)
\begin{theorem}[\cite{Weinstein1985}]
Let $d=1$. Then $\Null{L_1}=\mathbb{C}\sol'$.
\end{theorem}
\begin{proof}
The proof follows Weinstein \cite{Weinstein1985}.  We know one
solution, $\sol'$.  The fact that $\sol''(0)\neq 0$ 
allow us to choose the first linearly
independent solution to be $w_1 = \sol'/\sol''(0)$.  Then
$w_1(0)=0$, $w_1'(0)=1$.  Consider a second linearly independent
solution, $w_2$, with $w_2(0)=1$.  Since the Wronskian
\begin{equation}
   W[w_1,w_2] = w_1w_2'-w_1'w_2\; 
\end{equation}
is constant with respect to $x$, and since 
\begin{equation}
   W[w_1,w_2](x=0) = w_1(0) w'_2(0) - w_1'(0) w_2(0) = -w_2(0) = -1 \; ,
\end{equation}
we have $W[w_1,w_2]=-1$ for all $x$, and therefore 
\begin{equation}
   \sol'(x)w_2'(x) - \sol''(x)w_2(x) = -\sol''(0) >0 \; .
\end{equation}
The last equation can be rewritten as
\begin{equation}
   (\frac{w_2}{\sol'})' = \frac{-\sol''(0)}{(\sol')^2}.
\end{equation}
Now, for $\epsilon>0$ and $x>0$,
\begin{equation}
   w_2(x) = w_2(\epsilon)\frac{\sol'(x)}{\sol'(\epsilon)} 
        - \sol'(x)\sol''(0)\int_\epsilon^{x} 
        \frac{1}{(\sol'(\tilde{r}))^2} \diff \tilde{r} .
\end{equation}
Since $\sol'(x)<0$ for $x>0$, $\sol''(0)<0$,
and $|\sol'(x)|\leq C\exp{-c|x|}$, we have 
\begin{equation}
    w_2(x) \leq w_2(\epsilon)\frac{\sol'(x)}{\sol'(\epsilon)} - 
        \sol'(x)\sol''(0)\int_\epsilon^{x} \frac{\diff s}{C^2 \exp{-2cs}} \rightarrow -\infty 
        \ \ \text{as} \ \ x\rightarrow \infty \; .
\end{equation}
Hence $w_2\notin \Null{L_1}$ and we are done.\qed
\end{proof}
Next, we consider the case $d\geq 2$. We use the assumption that $\sol$
is spherically symmetric. In this case the operator $L_1$ (and also
$L_2$ and $\LL$) is spherically symmetric; \ie it commutes with the
action of the rotation group $\RotSpace(d)$ on $\Ltwo(\mathbb{R}^{d})$. 
As a result it can be
decomposed into a direct sum of ordinary differential operators
corresponding to the eigenfunction expansion of the Laplacian
$\Laplace_{\set{S}^{d-1}}$ on the sphere $\set{S}^{d-1}$. Denote the
orthonormal eigenfunctions of $-\Laplace_{\set{S}^{d-1}}$ 
corresponding to the eigenvalues
$\lambda_k=k(d-2+k)$ by $Y_k(\theta)$,
$\theta\in \set{S}^{d-1}$. Then the operator $L_1$ can we written as the
direct sum
\begin{equation}
   L_1 = \oplus A_{\freq,k}
\end{equation}
acting on the direct sum 
$\Ltwo(\mathbb{R}^d,\set{d}^{d}x)=\bigoplus_{k=0}^{\infty}
\Ltwo(\mathbb{R}_{+},r^{d-1}\set{d}r)\otimes Y_k(\theta)$, where the
operators $A_{\freq,k}$ are defined on $\Ltwo(\mathbb{R}_{+},r^{d-1}\set{d}r)$
by 
\begin{equation}
  A_{\freq,k} = -\Laplace_r + \freq + V_k(r),
\end{equation}
and where $\Laplace_r = \partial_r^2 + (d-1)r^{-1}\partial_r$ is the
radial Laplacian in $\mathbb{R}^d$ and
$V_k(r):=-f'(\sol)(r)+\lambda_k r^{-2}$. Clearly
$A_{\freq,k}^{*}=A_{\freq,k}$ and 
$\sigma_{\mathrm{ess}}(A_{\freq,k})=[\freq,\infty)$.

Now observe that 
\begin{equation}
   (\partial_{x_j}\sol)(x) = \hat{x}_j\sol'(r)\in
   \Ltwo(\mathbb{R}_{+},r^{d-1}\set{d}r)\otimes Y_1(\theta),
\end{equation}
where $\hat{x}=x|x|^{-1}\in \set{S}^{d-1}$. Hence
\begin{equation}
   \sol'\in \Null{A_{\freq,1}}.
\end{equation}
Since $\sol'(r)<0$ for $r>0$, we have by an extension of the
Perron-Frobenius (or Sturm-Liouville) theory (see \cite{OS1}) that
zero is the lowest eigenvalue of $A_{\freq,1}$ and is simple. Thus
\begin{equation}
A_{\freq,1} \geq 0\ \text{and}\ \Null{A_{\freq,1}}=\mathbb{C}\sol'.
\end{equation}

Next, since for $k\geq 2$
\begin{equation}
A_{\freq,k} - A_{\freq,1} = \frac{\lambda_k-\lambda_1}{r^2}>0,
\end{equation}
we conclude that $A_{\freq,k}>0$ for $k\geq 2$. 

\begin{theorem}
$\Null{A_{\freq,0}}$ is trivial provided $f(\psi)=h(|\psi|^{2})\psi$ 
with $h$ satisfying 
\begin{equation}\label{eq:mic}
    h'(r)+h''(r)r>0, \ r>0.
\end{equation}
\end{theorem}
\begin{proof}
    Assume there exists $\xi\in\Ltwo$ such that $\Am\xi = 0$. 
    Then $0$ is the second eigenvalue of $\Am$, and so the 
    corresponding eigenfunction $\xi$ has exactly one zero in 
    $(0,\infty)$, say at $r_{0}$.
Observe the following properties
\begin{enumerate}
    \item \label{pA1} $\xi\bot\Range{\Am}$;
    \item \label{pA2} $\Am \sol = 2h'(\sol^{2})\sol^{3}$;
    \item \label{pA3} $\Am \partial_{\freq}\sol = -\sol$.
\end{enumerate}
Properties~\ref{pA2} and~\ref{pA3} follow from the equation
\begin{equation}
    (-\Laplace_{r} + \freq)\sol - f(\sol) = 0
\end{equation}
for $\sol$.

Properties~\ref{pA1}--\ref{pA3} imply the relation
\begin{equation}\label{eq:sstar}
    \dotp{\xi}{(h'(\sol^{2})\sol^{2}-\alpha)\sol}=0, \ \forall 
    \alpha\in \mathbb{R}.
\end{equation}
Since $\sol$ is monotonically decreasing from some $\sol(0)>0$ at $0$ to 
$0$ as $r\rightarrow\infty$ (see \eg \cite{Berestycki+LionsI1983}),
and since $h'(s)s$ is monotonically 
increasing function of $s$ by condition \eqref{eq:mic}, we can choose 
$\alpha$ such that the monotonically decreasing function 
$h'((\sol(r))^2)\sol^{2}(r)-\alpha$ has a zero exactly at $r_{0}$.
In that case the left hand side of \eqref{eq:sstar} is non-zero, which 
leads to a contradiction. Thus the equation $\Am \xi = 0$ has no 
nontrivial solutions in $\Ltwo$. \qed
\end{proof}    

Finally we present the conditions under which McLeod's 
\cite{McLeod1993} uniqueness proof of positive solitons 
implies that $\Am$ has trivial null space in the case $d>1$.
There exists $\alpha>0$ such that
\begin{align}
   \freq s - sh(s^2)>0,&\ \text{for}\ 0<s<\alpha, \nonumber \\
   \freq s- sh(s^2)<0,&\ \text{for}\ \alpha<s<\infty,\\
   (sh(s^2)-\freq s)'>0,&\ \text{when}\ s=\alpha, \nonumber
\end{align}
and for each $S>\alpha$, $\exists
\lambda=\lambda(S)\in\set{C}((\alpha,\infty)$, $\mathbb{R}_+)$ such
that
\begin{equation}
  K(s,\lambda)\geq 0\ \text{for} \ s\in(0,S), \ \ 
   K(s,\lambda)\leq 0\ \text{for}\ s\in (S,\infty),  
\end{equation}
where
\begin{equation}
   K(s,\lambda)=\freq s + \lambda s^3 h'(s^2)
   -sh(s^2).  \nonumber
\end{equation}

\section{Coercivity of $\LL$}
\label{app:BBL}

The goal of this appendix is to prove the following result, essentially due to 
\cite{Weinstein1986}.
\begin{proposition}\label{prop:LL}
There is $\rho' > 0$ such that
if $w$ satisfies $\tfrm{w}{z}=0$ $\forall z\in \TM$, then 
\begin{equation}
\dotp{w}{\LL w}\geq \rho' \nrmHo{w}^2.
\end{equation}
\end{proposition}
\begin{proof}
We break the proof into three steps. The proof utilizes the fact that
$\LL$ has exactly one non-degenerate negative eigenvalue and the assumption
\ref{con:g.spec}; \ie that  
$\Null{\LL} = \Span\{(0 , \sol),
(\partial_{x_j} \sol, 0 ), j=1,...,d\}.
$

\begin{step}
   Let $\set{X}_1=\{w\in \Hone: \nrm{w}=1\; , \dotp{(\sol,0)}{w}=0\}$.
Then
\begin{equation}
   \inf_{w\in \set{X}_1} \rdot{w}{\LL w}=0 \; .\label{eq:L9}
\end{equation}
\end{step}
\begin{proof}
Let $\alpha:=\inf_{w\in \set{X}_1}\dotp{w}{\LL w}$, (see
\eqref{eq:Lm}).  Clearly $\nu\leq \alpha\leq 0$, where $\nu<0$ is the
negative eigenvalue of $\LL$ (see Appendix~\ref{app:spec}). That
$\alpha\leq 0$ is clear, as $w=(0,\sol)/\nrm{\sol}\in \set{X}_1$
yields $\dotp{w}{\LL w}=0$.  Moreover $\alpha\neq \nu$. Indeed if $\alpha=\nu$
then the minimizer, $v$, of \eqref{eq:L9} would be an 
eigenfunction of $\LL$ corresponding to the smallest
eigenvalue, $\nu$. 
Since
$(\sol,0)\bot v$, $(\sol,0)\bot \Null{\LL}$ and since $\nu$ is the
only negative eigenvalue of $\LL$ (see Proposition~\ref{prop:neg}), 
we conclude that $(\sol,0)$ is in the 
spectral subspace of $\LL$ corresponding to the interval
$[\delta,\infty)$ for some $\delta>0$. Therefore $\LL^{-1}(\sol,0)$ is
well defined and $\dotp{(\sol,0)}{\LL^{-1}(\sol,0)}>0$. On the other hand the
equation $\LL (\partial_\freq \sol,0)=-(\sol,0)$ implies that
\begin{equation}\label{eq:D3}
\dotp{(\sol,0)}{\LL^{-1}(\sol,0)}=-\mm'(\freq)<0
\end{equation}
by Condition \ref{con:g.stab} which contradicts 
$\dotp{(\sol,0)}{\LL(\sol,0)}>0$. Hence $\alpha=\nu$ is impossible.

To show that $\alpha=0$ we use the
Euler-Lagrange equations corresponding to \eqref{eq:L9}
\begin{equation}
\LL w = \alpha w + \beta (\sol,0)
\end{equation}
where $\alpha$ and $\beta$ are the Lagrange multipliers corresponding to 
$\nrm{w}=1$ and $\dotp{(\sol,0)}{w}=0$ respectively. Assume 
$\nu<\alpha<0$. 
If $\beta=0$, then
$\alpha$ would be a negative eigenvalue in $(\nu,0)$ which contradicts
that $\nu$ is the only negative eigenvalue. 
Thus $\beta\neq 0$.
Given $\nu<\alpha<0$, $\beta\neq 0$,
we can solve the Euler-Lagrange equation as
\begin{equation}
   w= \beta(\LL-\alpha)^{-1}(\sol,0).
\end{equation}
The inner product of the equation above with $(\sol,0)$, 
the orthogonality relation
$\rdot{w}{(\sol,0)}=0$, and $\beta\neq 0$, give
\begin{equation}
   0=\rdot{(\sol,0)}{(\LL+|\alpha|)^{-1}(\sol,0)}=:q(|\alpha|). \label{eq:g0}
\end{equation}
$q(\lambda)$ is analytic in $\lambda\in(0,|\nu|)$, 
and hence differentiable. Moreover, it
is monotonically decreasing, since 
\begin{equation}
   q'(\lambda) = -\rdot{(\sol,0)}{(\LL+\lambda)^{-2} (\sol,0)} =
   -\nrm{(\LL+\lambda)^{-1}(\sol,0)}^2<0.
\end{equation}
Furthermore by \eqref{eq:D3}
$q(0)=\rdot{(\sol,0)}{\LL^{-1}(\sol,0)} <0$.
Thus $q(|\alpha|)\neq 0$, for $\alpha\in (\nu,0)$, which contradicts 
\eqref{eq:g0}. Hence $\alpha=0$.\qed
\end{proof}
\begin{step}\label{lem:LLp}
   Let $\set{X}:=
\{w\in \Hone(\mathbb{R}^d,\mathbb{C}): \nrm{w}=1,\
\tfrm{w}{z}=0,\ \forall z\in \TM\}$. Then
\begin{equation}
   \inf_{w\in \set{X}} \rdot{w}{\LL w}>0 \; .\label{eq:LL9}
\end{equation}
\end{step}
\begin{proof}
The Euler-Lagrange equation corresponding to \eqref{eq:LL9} is
\begin{equation}
\LL w = \alpha w + \sum_k \gamma_k \J z_k
\end{equation} 
where $\{z_k\}$ is a basis for $\TM$. Here $\alpha$ and $\{\gamma_k\}$ are
the Lagrange multipliers corresponding to the constraints $\nrm{w}=1$ and
$\tfrm{w}{z_{k}}=0$ $\forall k$ respectively.
Note that $\alpha=\dotp{w}{\LL w}$, and that 
$\set{X}\subset \set{X}_1$, hence $\alpha\geq 0$. 
Assume that $\alpha=0$, 
and that one $\gamma_j\neq 0$.  Then for at least one  $z_k\in \TM$,
we have
\begin{equation}
\dotp{z_k}{\LL w} = \gamma_j (\om)_{jk} \neq 0,
\end{equation}
for some $k$, which contradicts $\dotp{z_{k}}{\LL w}=\dotp{\LL z_k}{w}=0$ 
$\forall k$. 
Here we have used that $\det\om\neq 0$, and that
$z_k$ is either a zero-eigenfunction or an associated zero-mode for
$\LL$.  Thus either $\alpha>0$ or $\gamma_j=0$. Consider the latter case. In this case
\begin{equation}
\LL w = 0.
\end{equation}
which implies that $w\in \Null{\LL}$. Since $\Null{\LL}\subset \TM$,
the relation $\tfrm{w}{z_k}=0$ for all $z_k\in
\TM$, contradicts the non-degeneracy of $\iom$ 
(see Corollary~\ref{cor:nondeg}).
Thus $\alpha>0$.\qed
\end{proof} \noindent
{\bf Step 3. End of Proof.}
\Eref{eq:LL9} implies that there exists a $\rho''>0$
such that
\begin{equation}
   \rdot{w}{\LL w}\geq \rho'' \nrm{w}^2\; , \label{eq:cL}
\end{equation}
for some $\rho''=\rho''(\freq)$.
To improve the coercivity from $\Ltwo$ to $\Hone$, we let $0<\delta<1$, and 
estimate $\dotp{w}{\LL w}$
using \eqref{eq:cL} as  
\begin{equation}
    (1-\delta)\rho''\nrm{w}^2 + 
        \delta \rdot{w}{\LL w}
    \leq 
        \rdot{w}{\LL w} \; .
\end{equation}
Upon using the explicit form of $\LL$ we find that
\begin{equation}
    \rdot{w}{\LL w} \geq
          \nrm{\nabla w}^2 - C_\freq\nrm{w}^2,
\end{equation}
where 
\begin{equation}
      C_\freq = \sup_x (\freq + |f'(\sol)|).
\end{equation}
The last two estimates with 
$\delta := \rho''(1+ \rho'' + C_\freq)^{-1}$ imply
\begin{equation}
   \rdot{w}{\LL w} \geq  \rho'\nrmHo{w}^2 \; ,
\end{equation}
where $\rho' = \rho''(1+\rho''+C_\freq)^{-1}$. 
This concludes the proof of Proposition~\ref{prop:LL}.\qed
\end{proof}

\bibliography{stein} 

\newcommand{\SortNoop}[1]{}
\begin{thebibliography}{10}

\bibitem{Adachi2002}
S.~Adachi.
\newblock A positive solution of a nonhomogeneous elliptic equation in
  $\mathbb{R}^{N}$ with {$G$}-invariant nonlinearity.
\newblock {\em Comm. PDE.}, 27(1\&2):1--22, 2002, doi:10.1081/PDE-120002781.

\bibitem{Arnold1989}
V.~I. Arnol'd.
\newblock {\em Mathematical methods of classical mechanics}.
\newblock Number~60 in Graduate Texts in Mathematics. Springer-Verlag, New
  York, second edition, 1989.

\bibitem{Berestycki+Gallouet+Kavian1983}
H.~Berestycki, T.~Gallouet, and O.~Kavian.
\newblock {\'E}quations de champs scalaires euclidiens non lin\'eaires dans le
  plan.
\newblock {\em C. R. Acad. Sci. Paris S\'er. I Math.}, 297(5):307--310, 1983.

\bibitem{Berestycki+LionsI1983}
H.~Berestycki and P.-L. Lions.
\newblock Nonlinear scalar field equations. {I}. {E}xistence of a ground state.
\newblock {\em Arch Rational Mech. Anal.}, 82(4):313--345, 1983.

\bibitem{Berestycki+LionsII1983}
H.~Berestycki and P.-L. Lions.
\newblock Nonlinear scalar field equations. {II. E}xistence of infinitely many
  solutions.
\newblock {\em Arch Rational Mech. Anal.}, 82(4):347--375, 1983.

\bibitem{Berestycki+Lions+Peletier1981}
H.~Berestycki, P.-L. Lions, and L.~A. Peletier.
\newblock An {ODE} approach to the existence of positive solutions for
  semilinear problems in {$\mathbb{R}^{N}$}.
\newblock {\em Indiana Univ. Math. J.}, 30(1):141--157, 1981.

\bibitem{Bronski+Jerrard2000}
J.~C. Bronski and R.~L. Jerrard.
\newblock Soliton dynamics in a potential.
\newblock {\em Math. Res. Lett.}, 7(2-3):329--342, 2000.

\bibitem{BP92}
V.~S. Buslaev and G.~S. Perel'man.
\newblock Scattering for the nonlinear {S}chr{\"o}dinger equation: states that
  are close to a soliton.
\newblock {\em Algebra i Analiz}, 4(6):63--102, 1992.

\bibitem{Buslaev+Perelman1995}
V.~S. Buslaev and G.~S. Perel'man.
\newblock On the stability of solitary waves for nonlinear {S}chr\"{o}dinger
  equations.
\newblock {\em Amer. Math. Soc. Transl. Ser.}, 2(164):74--98, 1995.

\bibitem{Buslaev+Sulem2002}
V.~S. Buslaev and C.~Sulem.
\newblock On asymptotic stability of solitary waves for nonlinear
  {S}chr{\"o}dinger equations.
\newblock {\em Ann. IHP. Analyse Nonlin\'eaire}, 20:419--475, 2003,
  doi:10.1016/S0294-1449(02)00018-5.

\bibitem{Cazenave1996}
T.~Cazenave.
\newblock {\em An introduction to nonlinear {S}chr\"{o}dinger equations}.
\newblock Number~26 in Textos de M\'etodos Matem\'aticos. Instituto de
  Matematica - UFRJ, Rio de Janeiro, RJ, third edition, 1996.

\bibitem{Cazenave+Lions82}
T.~Cazenave and P.-L. Lions.
\newblock Orbital stability of standing waves for some nonlinear
  {S}chr{\"o}dinger equations.
\newblock {\em Comm. Math. Phys.}, 85(4):549--561, 1982.

\bibitem{Cuccagna2001}
S.~Cuccagna.
\newblock Stabilization of solutions to nonlinear {S}chr\"{o}dinger equations.
\newblock {\em Comm. Pure Appl. Math.}, 54(9):1110--1145, 2001,
  doi:10.1002/cpa.1018.

\bibitem{Cuccagna2002}
S.~Cuccagna.
\newblock Asymptotic stability of the ground states of the nonlinear
  {S}chr{\"o}dinger equation.
\newblock {\em Rend. Istit. Mat. Univ. Trieste}, 32(suppl. 1):105--118, 2002.

\bibitem{Derks+Groesen}
G.~Derks and E.~van Groesen.
\newblock Energy propagation in dissipative systems. {P}art {II}:
  {C}entrovelocity for nonlinear wave equations.
\newblock {\em Wave Motion}, 15:159--172, 1992.

\bibitem{Frohlich+Tsai+Yau2000}
J.~Fr\"{o}hlich, T.-P. Tsai, and H.-T. Yau.
\newblock On a classical limit of quantum theory and the non-linear {H}artree
  equation.
\newblock {\em Geom. Funct. Anal.}, Special Volume, Part I:57--78, 2000.

\bibitem{Frohlich+Tsai+Yau2002}
J.~Fr\"{o}hlich, T.-P. Tsai, and H.-T. Yau.
\newblock On the point-particle ({N}ewtonian) limit of the non-linear {H}artree
  equation.
\newblock {\em Comm. Math. Phys.}, 225(2):223--274, 2002,
  doi:10.1007/s002200100579.

\bibitem{Ginibre+Velo1979}
J.~Ginibre and G.~Velo.
\newblock On a class of nonlinear {S}chr\"{o}dinger equations. {I,II}.
\newblock {\em J. Func. Anal.}, 32:1--71, 1979.

\bibitem{Ginibre+Velo1980}
J.~Ginibre and G.~Velo.
\newblock On a class of nonlinear {S}chr{\"o}dinger equation with nonlocal
  interaction.
\newblock {\em Math. Z.}, 170(2):109--136, 1980.

\bibitem{Grillakis+Shatah+Strauss1987}
M.~Grillakis, H.~Shatah, and W.~Strauss.
\newblock Stability theory of solitary waves in the presence of symmetry. {I}.
\newblock {\em J. Funct. Anal.}, 74(1):160--197, 1987.

\bibitem{Grillakis+Shatah+Strauss1990}
M.~Grillakis, H.~Shatah, and W.~Strauss.
\newblock Stability theory of solitary waves in the presence of symmetry. {II}.
\newblock {\em J. Funct. Anal.}, 94(2):308--348, 1990.

\bibitem{Groesen+Mainardi}
E.~S. Groesen and F.~Mainardi.
\newblock Energy propagation in dissipative systems. {P}art~{I}:
  {C}entrovelocity for linear systems.
\newblock {\em Wave Motion}, 11:201--209, 1989.

\bibitem{GS}
S.~Gustafson and I.~M. Sigal.
\newblock Dynamics of magnetic vortices.
\newblock preprint, 2003.

\bibitem{Jones+Kupper1986}
C.~K. R.~T. Jones and T.~K\"{u}pper.
\newblock On the infinitely many solutions of a semilinear elliptic equation.
\newblock {\em SIAM J. Math. Anal.}, 17(4):803--835, 1986.

\bibitem{Kato1987}
T.~Kato.
\newblock On nonlinear {S}chr\"{o}dinger equations.
\newblock {\em Ann. IHP. Physique Th\'eorique}, 46:113--129, 1987.

\bibitem{Li+Li1993}
C.~Li and Y.~Y. Li.
\newblock Nonautonomous nonlinear scalar field equations in {$\mathbb{R}\sp
  2$}.
\newblock {\em J. Diff. Eqn.}, 103(2):421--436, 1993.

\bibitem{Li1990}
Y.~Y. Li.
\newblock Nonautonomous nonlinear scalar field equations.
\newblock {\em Indiana Univ. Math. J.}, 39(2):283--301, 1990.

\bibitem{Lions1986b}
P.-L. Lions.
\newblock On positive solutions of semilinear elliptic equations in unbounded
  domains. nonlinear diffusion equations and their equilibrium states, {II}
  ({B}erkeley, {CA, 1986}.
\newblock {\em Math. Sci. Res. Inst. Publ.}, 13:85--122, 1988.

\bibitem{McLeod1993}
K.~Mc{L}eod.
\newblock Uniqueness of positive radial solutions of {$\Laplace u + f(u)=0$} in
  {$\mathbb{R}^n$}, {II}.
\newblock {\em Am. Math. Soc.}, 339(2):495--505, 1993.

\bibitem{McLeod+Serrin1987}
K.~Mc{L}eod and J.~Serrin.
\newblock Uniqueness of positive radial solutions of {$\Laplace u + f(u)=0$} in
  {$\mathbb{R}^n$}.
\newblock {\em Arch. Rational Mech. Anal.}, 99(2):115--145, 1987.

\bibitem{OS1}
Y.~N. Ovchinnikov and I.~M. Sigal.
\newblock Ginzburg-{L}andau equation. {I}. {S}tatic vortices.
\newblock In {\em Partial differential equations and their applications
  (Toronto, ON, 1995)}, number~12 in CRM Proc. Lecture Notes, pages 199--220.
  Amer. Math. Soc., Providence, RI, 1997.

\bibitem{Peletier+Serrin1983}
L.~A. Peletier and J.~Serrin.
\newblock Uniqueness of positive solutions of semilinear equations in
  {$\mathbb{R}^{n}$}.
\newblock {\em Arch. Rational Mech. Anal.}, 81(2):181--197, 1983.

\bibitem{Pelinovsky1996}
D.~E. Pelinovsky, V.~V. Afanasjev, and Y.~S. Kivshar.
\newblock Nonlinear theory of oscillating, decaying, and collapsing solitons in
  the general nonlinear {S}chr{\"o}dinger equation.
\newblock {\em Phys. Rev. E}, 53(2):1940--53, 1996,
  doi:10.1103/PhysRevE.53.1940.

\bibitem{Pelinovsky+Grimshaw1997}
D.~E. Pelinovsky and R.~H.~J. Grimshaw.
\newblock Asymptotic methods in soliton stability theory.
\newblock In L.~Debnath and S.~R. Choudhury, editors, {\em Nonlinear
  instability analysis}, volume~12, pages 245--312. Comput. Mech., Southampton,
  1997.

\bibitem{Perelman2001}
G.~S. Perel'man.
\newblock Preprint, 2001.

\bibitem{Reed+Simon+4}
M.~Reed and B.~Simon.
\newblock {\em Methods of Modern Mathematical Physics. {IV A}nalysis of
  Operators}.
\newblock Academic Press, New York, 1978.

\bibitem{RSS}
I.~Rodnianski, W.~Schlag, and A.~Soffer.
\newblock Asymptotic stability of $n$-soliton states of {NLS}.
\newblock preprint, 2003, ArXiv:math.AP/0309114.

\bibitem{Soffer+Weinstein1990}
A.~Soffer\SortNoop{A} and M.~I. Weinstein.
\newblock Multichannel nonlinear scattering for nonintegrable equations.
\newblock {\em Comm. Math. Phys.}, 133(1):119--146, 1990.

\bibitem{Soffer+Weinstein1992}
A.~Soffer\SortNoop{B} and M.~I. Weinstein.
\newblock Multichannel nonlinear scattering for non{\-}integrable equations
  {II}. {T}he case of anisotropic potentials and data.
\newblock {\em J. Differential Equations}, 98(2):376--390, 1992.

\bibitem{Soffer+Weinstein2003}
A.~Soffer\SortNoop{C} and M.~I. Weinstein.
\newblock Selection of the ground state for nonlinear {S}chr\"odinger
  equations.
\newblock Preprint, 2003, ArXiv:nlin.PS/0308020.

\bibitem{Strauss1977}
W.~A. Strauss.
\newblock Existence of solitary waves in higher dimensions.
\newblock {\em Comm. Math. Phys.}, 55(2):149--162, 1977.

\bibitem{Stuart2001}
D.~M.~A. Stuart.
\newblock Modulation approach to stability of non-topological solitions in
  semilinear wave equations.
\newblock {\em J. Math. Pures Appl.}, 80(1):51--83, 2001,
  doi:10.1016/S0021-7824(00)01189-2.

\bibitem{Sulem1999}
C.~Sulem and P.-L. Sulem.
\newblock {\em The Nonlinear {S}chr{\"{o}}dinger Equation. Self-Focusing and
  Wave Collapse}.
\newblock Number 139 in Applied Mathematical Sciences. Springer, New York,
  1999.

\bibitem{Tsai+Yau2002}
T.-P. Tsai and H.-T. Yau.
\newblock Asymptotic dynamics of nonlinear {S}chr\"{o}dinger equations:
  resonance-dominated and dispersion-dominated solutions.
\newblock {\em Comm. Pure Appl. Math.}, 55(2):153--216, 2002,
  doi:10.1002/cpa.3012.

\bibitem{Tsai+Yau2002b}
T.-P. Tsai and H.-T. Yau.
\newblock Relaxation of excited states in nonlinear {S}chr\"odinger equations.
\newblock {\em Int. Math. Res. Not.}, 2002(31):1629--1673, 2002,
  doi:10.1155/S1073792802201063.

\bibitem{Tsai+Yau2002c}
T.-P. Tsai and H.-T. Yau.
\newblock Stable directions for excited states of nonlinear {S}chr{\"o}dinger
  equations.
\newblock {\em Comm. PDE}, 27(11-12):2363--2402, 2002,
  doi:10.1081/PDE-120016161.

\bibitem{Weinstein1985}
M.~I. Weinstein.
\newblock Modulational stability of ground states of nonlinear
  {S}chr\"{o}dinger equations.
\newblock {\em SIAM J. Math. Anal.}, 16(3):472--491, 1985.

\bibitem{Weinstein1986}
M.~I. Weinstein.
\newblock Lyapunov stability of ground states of nonlinear dispersive evolution
  equations.
\newblock {\em Comm. Pure Appl. Math.}, XXXIX:51--68, 1986.

\end{thebibliography}
\bibliographystyle{habbrv}

\end{document}